\documentclass[preprintnumbers,secnumarabic]{revtex4}

\usepackage{graphicx}
\usepackage{amsmath,amssymb,mathrsfs}
\usepackage{fancyhdr}
\usepackage{psfrag}
\usepackage{array,bbm} 
\usepackage{url}
\usepackage{subfig}

\newcommand*{\CPb}{\ensuremath{\text{CP}_{g,2}^{(ii)}}}

\newcommand*{\trans}{\mathrm{T}}                     

\newcommand*{\tvec}[1]{\ensuremath{\boldsymbol{\mathrm{#1}}}}           
 
\makeatletter       

\renewcommand{\p@subsection}{}

\makeatother

\begin{document}

\preprint{HD-THEP-09-25}

\title{On the phenomenology of a two-Higgs-doublet model with
maximal CP symmetry at the LHC\\
II: radiative effects}

\author{M. Maniatis}
    \email[E-mail: ]{M.Maniatis@thphys.uni-heidelberg.de}
\author{O. Nachtmann}
    \email[E-mail: ]{O.Nachtmann@thphys.uni-heidelberg.de}


\affiliation{
Institut f\"ur Theoretische Physik, Philosophenweg 16, 69120
Heidelberg, Germany
}

\begin{abstract}
The processes $p \bar{p} \rightarrow \gamma + \text{heavy-flavour jet(s)} + X$
and
$p p \rightarrow \gamma + \text{heavy-flavour jet(s)} + X$ are
studied in the framework of a special two-Higgs-doublet model, the MCPM.
As distinguishing feature of this model we find that radiative Higgs-boson
production and decay lead to heavy flavour $c$ jets but no $b$ jets in the
above processes. 
Thus, the prediction is that $b$ jets should be given by the normal QCD
processes whereas for $c$ jets an excess over the QCD expectation should occur.
We present results both for Tevatron and LHC energies.
\end{abstract}

\maketitle

%
\section{Introduction}
\label{sec-intro}

Two-Higgs-doublet models (THDMs) have been studied extensively since many years (see
\cite{Kobayashi:1973fv,Lee:1973iz,Lee:1974jb,Glashow:1976nt,Paschos:1976ay,Gunion:1989we,
Cvetic:1993cy,Ginzburg:2004vp,Gunion:2005ja,
Barbieri:2005kf,Branco:2005em,Barbieri:2006dq,Nishi:2006tg,
Ivanov:2006yq,Fromme:2006cm,Barroso:2007rr,Gerard:2007kn,Ma:2009ax,
Ferreira:2009wh,Mahmoudi:2009zx} and references
therein).
These models provide a simple extension of the Standard Model of particle
physics (SM) with a rich phenomenology due to the extended Higgs sector.
In our group we have studied various aspects of the general multi-Higgs
models and THDMs in particular~\cite{Bernreuther:1998rx,Nagel:2004sw,Maniatis:2006fs,Maniatis:2006jd}.
A special two-Higgs doublet model, the {\em maximally-CP-symmetric model}
(MCPM), was introduced in~\cite{Maniatis:2007vn,Maniatis:2007de}.
The phenomenology of the MCPM for high-energy proton--antiproton
and proton--proton collisions was developed in detail in~\cite{Maniatis:2009vp}.
In the present article we continue the discussion of the MCPM
phenomenology in view of processes involving real photons plus heavy
flavour jets in the final state. That is, we consider the processes
\begin{align}
\label{eq1}
p + \bar{p} \longrightarrow&\; \gamma + \text{heavy-flavour jet(s)} +X\,,\\
\intertext{relevant for the Tevatron physics, and}
\label{eq2}
p + p \longrightarrow&\; \gamma + \text{heavy-flavour jet(s)} +X
\end{align}
relevant for the LHC.

The general motivation for the MCPM is summarised in~\cite{Maniatis:2009vp}.
This will not be repeated her since the present paper is a direct
continuation of this work. 
Let us recall from~\cite{Maniatis:2009vp} that in the MCPM we
have -- as in every THDM -- five physical Higgs bosons, three neutral
ones and a charged pair:
\begin{equation}
\rho',\quad h', \quad h'', \quad H^\pm\;.
\end{equation}
Distinguishing features of the MCPM as discussed in~\cite{Maniatis:2009vp}
are as follows.
\begin{itemize}
\item The Higgs bosons $h'$, $h''$, and $H^\pm$ couple exclusively
to the second fermion family~($\nu_\mu$, $\mu$, $c$, $s$),
but with coupling constants proportional to the masses of the 
third fermion family~($\nu_\tau$, $\tau$, $t$, $b$).
\item For Higgs-boson masses below about 300~GeV the $\rho'$ behaves
essentially as the Higgs boson $\rho'_{\text{SM}}$ of the SM.
\item The main decays of the other Higgs bosons are
\begin{equation}
h'  \longrightarrow \; c \bar{c}\,,\qquad
h'' \longrightarrow \; c \bar{c}\,,\qquad
H^+ \longrightarrow \; c \bar{s}\,,\qquad
H^- \longrightarrow \; s \bar{c}\,.
\end{equation}
\item The main production modes of these Higgs bosons
in high energy $p \bar{p}$ and $p p$ collisions
are Drell--Yan-type processes,
that is, quark--antiquark fusion giving a Higgs boson. 
\begin{figure}[h]
\subfloat[]{
\includegraphics[height=0.2\textwidth,clip]{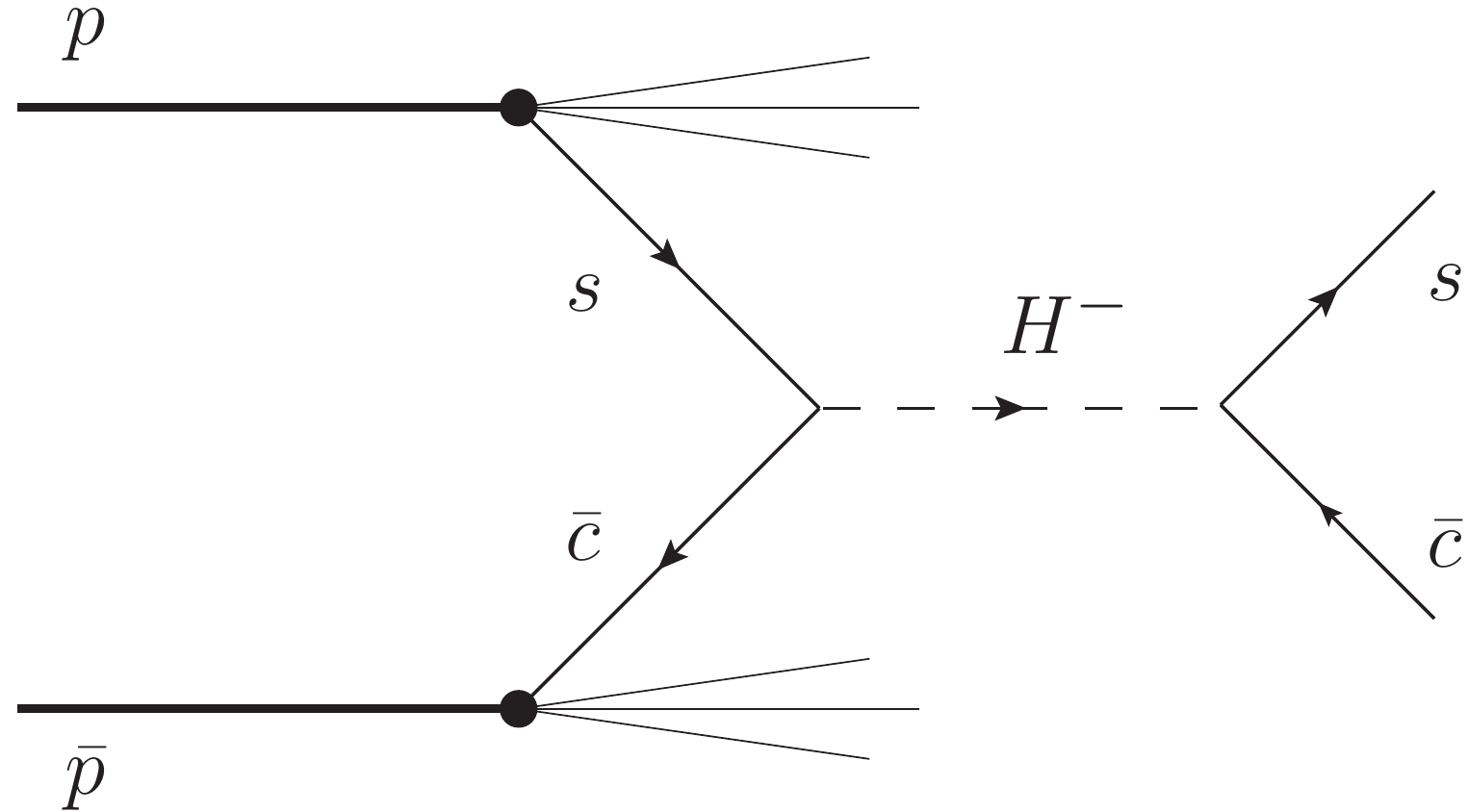}
\label{figppbHsc}
 }\qquad
\subfloat[]{
\includegraphics[height=0.2\textwidth,clip]{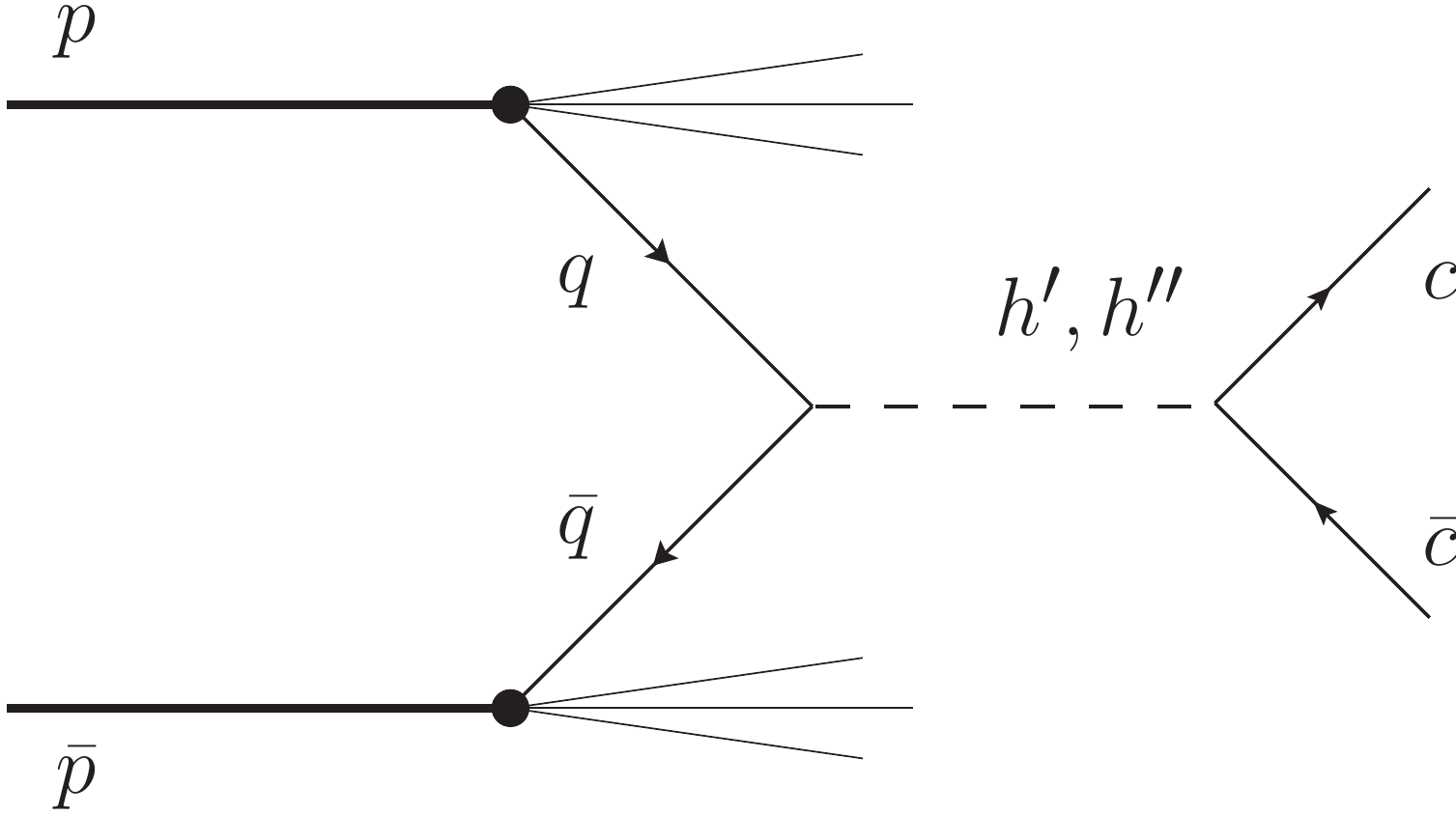}
\label{figpphpcc}
}
\caption{(a): Drell--Yan process for
production of a Higgs-boson $H^-$ by $s \bar{c}$ fusion followed
by the decay $H^- \longrightarrow \; s \bar{c}$.
(b):
Drell--Yan process for
production of the Higgs-bosons $h'$ and $h''$ with
$q \bar{q}$ fusion where $q=s$ and $c$. This
is followed by the decays 
$h' \longrightarrow \; c \bar{c}$ and
$h'' \longrightarrow \; c \bar{c}$, respectively.
}
\end{figure}
\end{itemize}
%

In Fig.\ref{figppbHsc} we show the diagram for the Drell--Yan production
of the $H^-$ followed by its main decay for $p \bar{p}$ collisions. For
$p p$ collisions one just has to replace $\bar{p}$ by $p$.
For the $H^+$ production and decay the replacements $s \rightarrow \bar{s}$
and $\bar{c} \rightarrow c$ have to be made. In Fig.~\ref{figpphpcc}
the Drell--Yan production processes of $h'$ and $h''$ are shown followed by their main
decays. The cross sections for these processes were calculated in~\cite{Maniatis:2009vp}.
For Higgs-boson masses of 100 to 300~GeV we find cross sections of 
order 300 to 1~pb for $p \bar{p}$ collisions at c.m energy 1.96~TeV. For
$p p$ collisions at c.m. energy 14~TeV these cross sections are of order
5000 to 500~pb. All these are relatively large cross sections.

The aim of the present article is to calculate the processes as shown in
Figs.~\ref{figppbHsc} and \ref{figpphpcc} but with the additional emission of a 
real photon. Clearly, this will lead to a photon plus one or two charm-quark
jets in the final state. The distinguishing feature of the MCPM is that,
as far as heavy quark flavours are concerned, only charm-quark jets are produced
by this mechanism, no bottom-quark jets.

Our paper is organised as follows.
In Sect.~\ref{secII} we discuss the radiative production
and the radiative decay of $H^\pm$. The analogous processes
for the neutral Higgs bosons $h'$ and $h''$ are studied
in Sect.~\ref{secIII}. In Sect.~\ref{secIV} we present
numerical results. 
In Sect.~\ref{secV} we discuss the possible relevance of our
results in view of experimental results~\cite{:2009de} from the
Tevatron and we present our conclusions.
The appendix contains details of our 
calculations. All our notations and kinematic conventions are
the same as in~\cite{Maniatis:2009vp} and follow~\cite{Nachtmann:1990ta}.

%
\section{Radiative production of $H^\pm$ and $H^\pm$ radiative decay}
\label{secII}

\begin{figure}
\subfloat[]{\includegraphics[width=.25\linewidth]{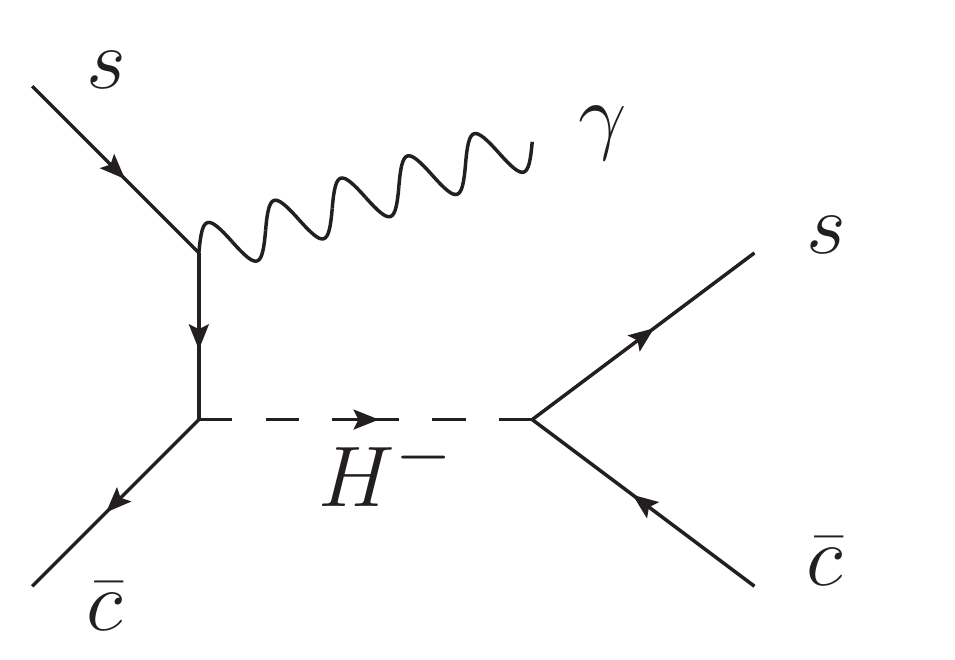}}
\subfloat[]{\includegraphics[width=.25\linewidth]{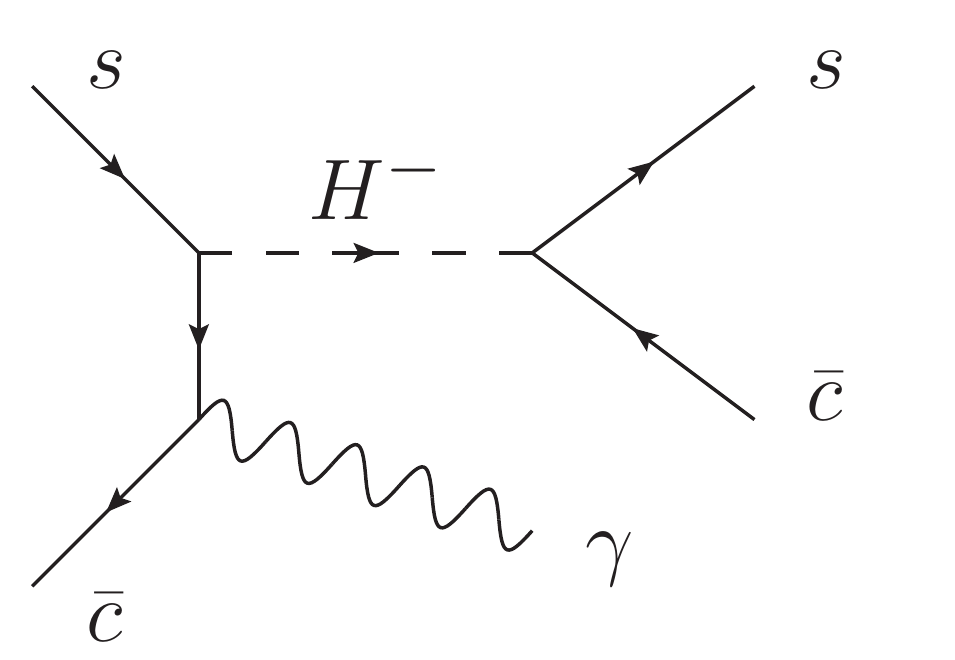}}
\subfloat[]{\includegraphics[width=.25\linewidth]{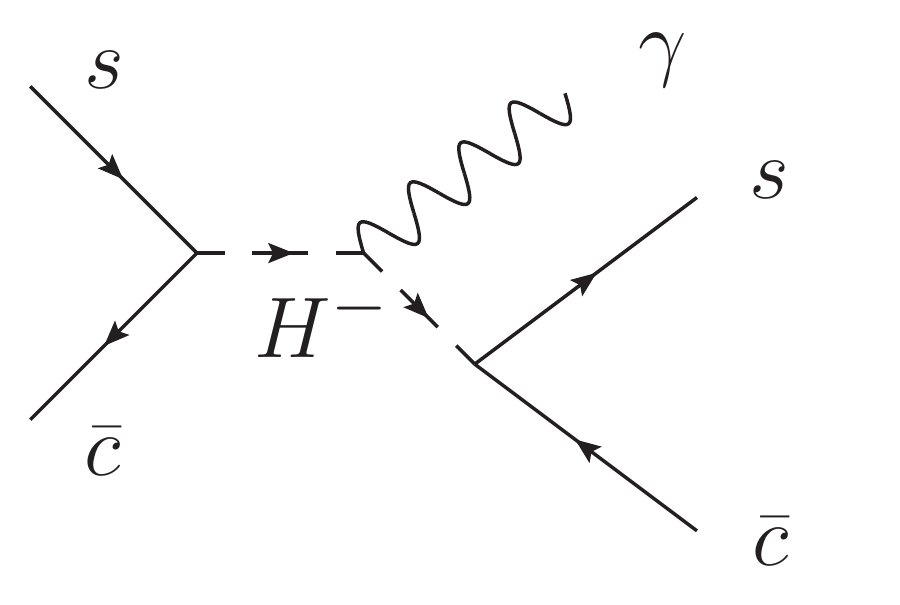}}\\
\subfloat[]{\includegraphics[width=.25\linewidth]{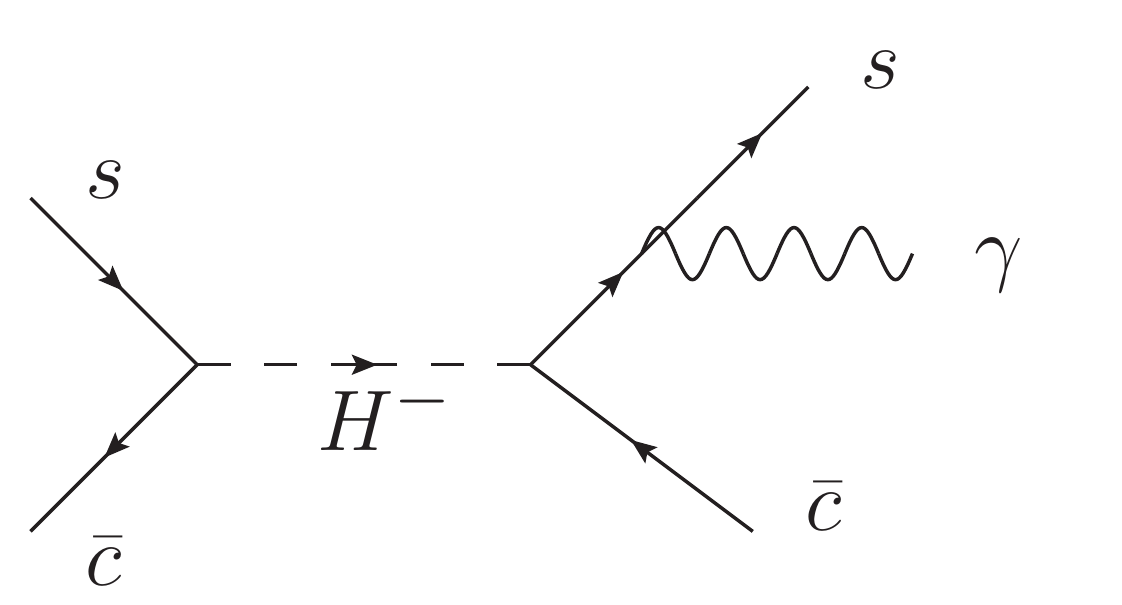}}
\subfloat[]{\includegraphics[width=.25\linewidth]{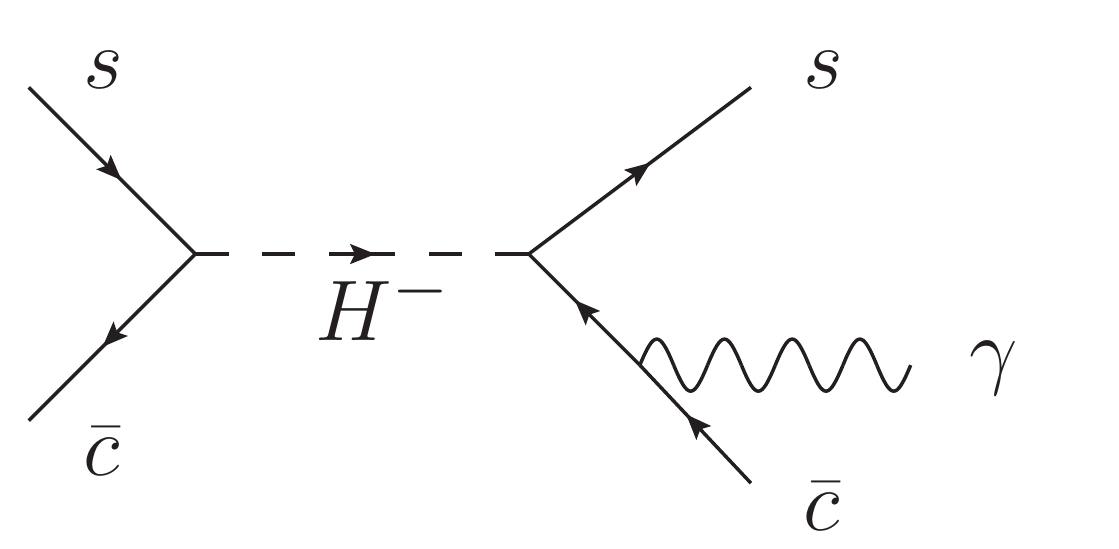}}
\caption{\label{fig3}The five diagrams for real photon emission in
the basic Drell--Yan process of $H^-$ production and decay
of Fig.~\ref{figppbHsc}.}
\end{figure}

In this section we discuss first the real photon emission in $H^-$
production and decay as shown in Fig.~\ref{figppbHsc}. Treating the
$s$ and $\bar{c}$ quarks as on-shell particles we have in lowest
order of the electromagnetic coupling $e = \sqrt{4 \pi \alpha_{\text{em}}}$
five possibilities to attach a photon line to the basic diagram
of Fig.~\ref{figppbHsc}. This is shown in Fig.~\ref{fig3}.
In the following we shall always work in the
narrow width approximation for the Higgs bosons. Then the
diagrams of Fig.~\ref{fig3} 
can be considered as representing two distinct processes.
We have firstly the radiative production of 
a {\em real} $H^-$
\begin{equation}
\label{eq5}
s(p_1') + \bar{c}(p_2') \longrightarrow\; H^-(k_1) + \gamma(k_2)\,.
\end{equation}
The $H^-$ decays then to a $s \bar{c}$ quark pair $H^- \longrightarrow\; s \bar{c}$.
The diagrams (a), (b), and (c) of Fig.~\ref{fig3} contribute to \eqref{eq5}.

Secondly, we have the Drell--Yan production of a {\em real} $H^-$
 which then decays to $s \bar{c} \gamma$:
\begin{equation}
\label{eq6}
 H^-(k_1) \longrightarrow\; s(p_1') + \bar{c}(p_2') + \gamma(k_2)\,.
\end{equation}
To \eqref{eq6} the diagrams (c), (d), and (e)
of Fig.~\ref{fig3} contribute.

In this section we shall give the results for the processes~\eqref{eq5} 
and \eqref{eq6}. The details of the calculation are given
in appendix~\ref{appA}.

\subsection{Radiative $H^\pm$ production}

The transition rate for radiative $H^-$ production~\eqref{eq5}
is easily calculated. We get
\begin{multline}
\label{eq7}
d \Gamma \big( s(p_1') + \bar{c}(p_2') \rightarrow\; H^-(k_1) + \gamma(k_2) \big)
= \frac{1}{V} \frac{1}{2 p_1'^0 2 p_2'^0}
\frac{1}{(2 \pi)^2} \frac{1}{4 N_c^2}\\
\times\;I_{s \bar{c}}^{H^-} (p_1', p_2', k_1, k_2)
\delta^{(4)}(p_1'+p_2'-k_1-k_2)
\frac{d^3 k_1}{2 k_1^0}
\frac{d^3 k_2}{2 k_2^0}\,,
\end{multline}
where $N_c=3$ is the number of colours, $V$ is the normalisation
volume, and the function $I_{s \bar{c}}^{H^-}$
is given in appendix~\ref{appA}. Using~\eqref{eq7} we can, in the
usual way, see for instance~\cite{Nachtmann:1990ta}, 
calculate the Drell--Yan production of $H^- + \gamma$
in $p \bar{p}$ collisions
\begin{equation}
\label{eq8}
p(p_1) + \bar{p}(p_2) \longrightarrow\; H^-(k_1) + \gamma(k_2) + X\,.
\end{equation}
We get for the cross section (for $p \bar{p}$ c.m. energy squared
$s=(p_1+p_2)^2 \gg m_p^2)$ the following
\begin{multline}
\label{eq10}
d \sigma \big( p(p_1) + \bar{p}(p_2) \longrightarrow\; H^-(k_1) + \gamma(k_2) + X \big) =
\frac{1}{2 s}
\frac{1}{4 N_c^2}
\frac{1}{(2 \pi)^2}\;\\
\times\intop_0^1 d x_1
\intop_0^1 d x_2
\bigg\{ 
\frac{1}{x_1} N_s^p(x_1) 
\frac{1}{x_2} N_{\bar{c}}^{\bar{p}}(x_2) 
I_{s \bar{c}}^{H^-} (x_1 p_1, x_2 p_2, k_1, k_2)
+
\frac{1}{x_1} N_{\bar{c}}^p(x_1) 
\frac{1}{x_2} N_{s}^{\bar{p}}(x_2) 
I_{s \bar{c}}^{H^-} (x_2 p_2, x_1 p_1, k_1, k_2) \bigg\}\\
\times\delta^{(4)}(x_1 p_1+x_2 p_2-k_1-k_2)
\frac{d^3 k_1}{2 k_1^0}
\frac{d^3 k_2}{2 k_2^0}\,.
\end{multline}
Here $N_q^p(x)$ and $N_q^{\bar{p}}(x)$ are the usual
parton distribution functions (pdf's) at the
c.m. energy scale $\sqrt{s}$ considered.

In a similar way we can calculate the cross section for the
$H^+$ production in $p \bar{p}$ collisions
\begin{equation}
\label{eq11}
p(p_1) + \bar{p}(p_2) \longrightarrow\; H^+(k_1) + \gamma(k_2) + X\,.
\end{equation}
Here the basic process is
\begin{equation}
\label{eq12}
\bar{s}(p_1') + c(p_2') \longrightarrow\; H^+(k_1) + \gamma(k_2)\,.
\end{equation}
The cross section for~\eqref{eq11} is given by
\begin{multline}
\label{eq13}
d \sigma \big( p(p_1) + \bar{p}(p_2) \longrightarrow\; H^+(k_1) + \gamma(k_2) + X \big) =
\frac{1}{2 s}
\frac{1}{4 N_c^2}
\frac{1}{(2 \pi)^2}\;\\
\times\intop_0^1 d x_1
\intop_0^1 d x_2
\bigg\{
\frac{1}{x_1} N_{\bar{s}}^p(x_1) 
\frac{1}{x_2} N_c^{\bar{p}}(x_2) 
I_{\bar{s} c}^{H^+} (x_1 p_1, x_2 p_2, k_1, k_2)
+
\frac{1}{x_1} N_{c}^p(x_1) 
\frac{1}{x_2} N_{\bar{s}}^{\bar{p}}(x_2) 
I_{\bar{s} c}^{H^+} (x_2 p_2, x_1 p_1, k_1, k_2) \bigg\}\\
\times\delta^{(4)}(x_1 p_1+x_2 p_2-k_1-k_2)
\frac{d^3 k_1}{2 k_1^0}
\frac{d^3 k_2}{2 k_2^0}\,,
\end{multline}
where, as shown in appendix~\ref{appA},
\begin{equation}
\label{eq14}
I_{\bar{s} c}^{H^+} ( p_1', p_2', k_1, k_2)=
I_{s \bar{c}}^{H^-} (p_1', p_2', k_1, k_2)\,.
\end{equation}

For the processes
\begin{equation}
\label{eq15}
p(p_1) + p(p_2) \longrightarrow\; H^\mp(k_1) + \gamma(k_2) + X
\end{equation}
the cross sections read as in~\eqref{eq10} and \eqref{eq13}
with the replacements
\begin{equation}
N_{\bar{q}}^{\bar{p}} \longrightarrow N_{\bar{q}}^p, \qquad
N_q^{\bar{p}} \longrightarrow N_q^p\,,
\end{equation}
where $q=s, c$.

In Sect.~\ref{secIV} we shall give numerical results for the 
distributions in the transverse photon momentum $p_\trans^\gamma=|{\tvec{k}_2}_\trans|$
for the processes considered above.

\subsection{Radiative decays of $H^\mp$}

Here we study the decays
\begin{equation}
\label{eq17}
H^-(k_1) \longrightarrow\;s(p_1') + \bar{c}(p_2')  + \gamma(k_2)
\end{equation}
and
\begin{equation}
\label{eq18}
H^+(k_1) \longrightarrow\; \bar{s}(p_1') + c(p_2')  + \gamma(k_2)\;.
\end{equation}
The calculations presented in appendix~\ref{appA} give for
the differential decay rates
\begin{multline}
\label{eq19}
d \Gamma \big(  H^-(k_1) \rightarrow\;s(p_1') + \bar{c}(p_2')  + \gamma(k_2) \big) =
\frac{1}{2 m_{H^-}} 
\frac{1}{(2 \pi)^5} 
\;\tilde{I}_{s \bar{c}}^{H^-} (p_1', p_2', k_1, k_2) \\
\times \delta^{(4)}(p_1'+p_2'+k_2-k_1)
\frac{d^3 p_1'}{2 p_1'^0}
\frac{d^3 p_2'}{2 p_2'^0}
\frac{d^3 k_2}{2 k_2^0}\;,
\end{multline}

\begin{multline}
\label{eq20}
d \Gamma \big(  H^+(k_1) \rightarrow\;\bar{s}(p_1') + c(p_2')  + \gamma(k_2) \big) =
\frac{1}{2 m_{H^+}} 
\frac{1}{(2 \pi)^5} 
\;\tilde{I}_{\bar{s} c}^{H^+} (p_1', p_2', k_1, k_2) \\
\times\delta^{(4)}(p_1'+p_2'+k_2-k_1)
\frac{d^3 p_1'}{2 p_1'^0}
\frac{d^3 p_2'}{2 p_2'^0}
\frac{d^3 k_2}{2 k_2^0}\;.
\end{multline}
Here the CP and crossing symmetry relations 
give (see appendix~\ref{appA})
\begin{equation}
\label{eq21}
\tilde{I}_{s \bar{c}}^{H^-} (p_1', p_2', k_1, k_2)=
\tilde{I}_{\bar{s} c}^{H^+} (p_1', p_2', k_1, k_2)=
I_{s \bar{c}}^{H^-} (p_1', p_2', k_1, -k_2)
\end{equation}
with $k_1=p_1'+p_2'+k_2$ due to energy-momentum conservation.

In Sect.~\ref{secIV} we shall use~\eqref{eq19} and \eqref{eq20} to get
the contribution to the $\gamma$ flux from the ordinary Drell--Yan
production of $H^-$ and $H^+$ followed by the radiative decays~\eqref{eq17}
and \eqref{eq18}, respectively.

%
\section{Radiative production and radiative decay of $h'$ and $h''$}
\label{secIII}

\begin{figure}
\subfloat[]{\includegraphics[width=.25\linewidth]{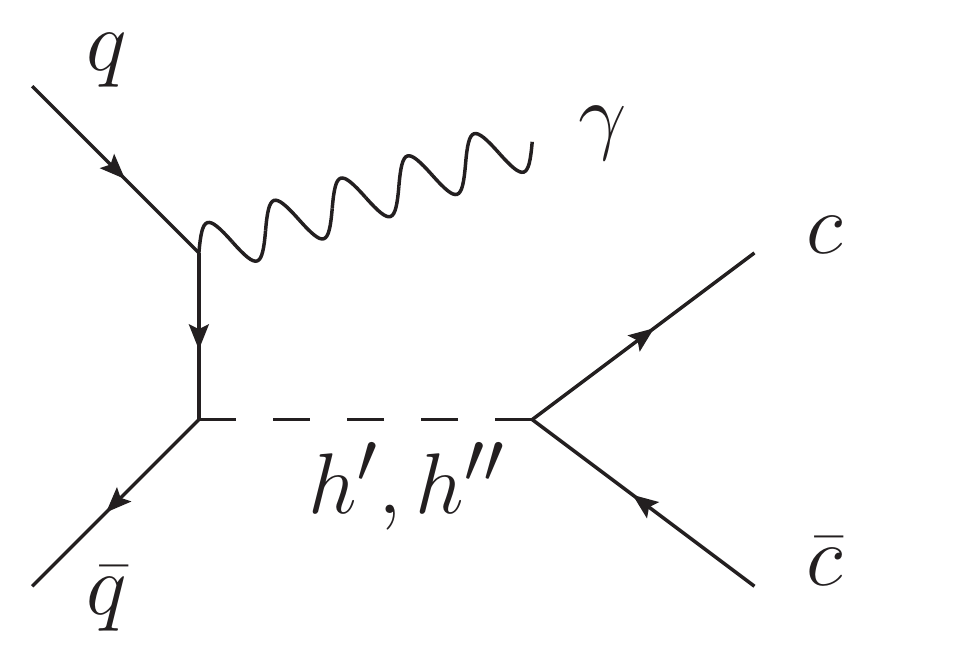}}
\subfloat[]{\includegraphics[width=.25\linewidth]{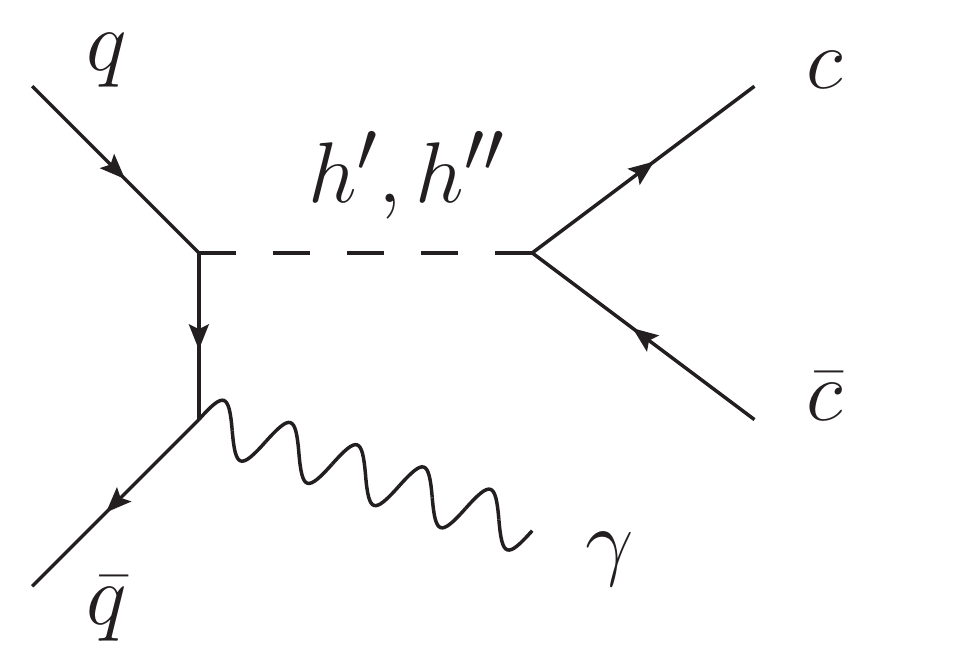}}
\subfloat[]{\includegraphics[width=.25\linewidth]{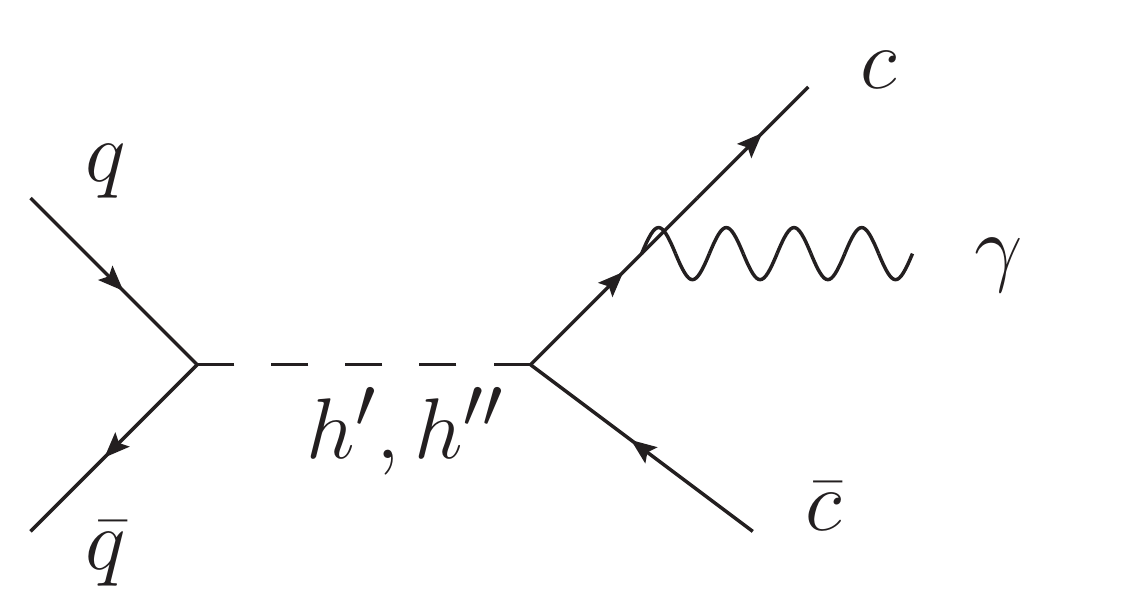}}
\subfloat[]{\includegraphics[width=.25\linewidth]{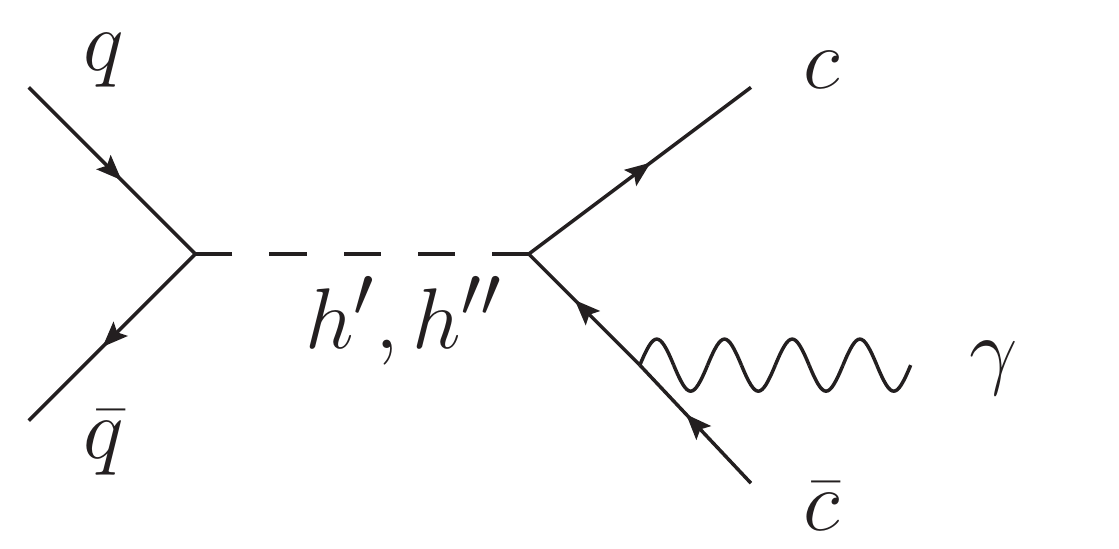}}
\caption{\label{fig4}The four diagrams for real photon emission in
the basic Drell--Yan process for $h'$ and $h''$ production and decay
of Fig.~\ref{figpphpcc}. Here $q=s, c$.}
\end{figure}

The Drell--Yan production of $h'$ and $h''$ followed by the main
decay of these bosons is shown in Fig.~\ref{figpphpcc}. The corresponding
diagrams for the process with emission of a real photon in addition
are shown in Fig.~\ref{fig4}.

We use the narrow width approximation for $h'$ and $h''$
as for $H^\pm$. The diagrams (a) and (b) of Fig.~\ref{fig4}
correspond then to the production of a {\em real} $h'$ and $h''$
which subsequently decays to a $c \bar{c}$ pair
\begin{equation}
\label{eq22}
q(p_1') + \bar{q}(p_2') \longrightarrow\; h(k_1) + \gamma(k_2)\,.
\end{equation}
Here and in the following we write
generically $h$ for $h'$ and $h''$.
The diagrams (c) and (d) of Fig.~\ref{fig4} correspond to the
ordinary Drell--Yan production of $h$ followed
by the radiative decay
\begin{equation}
\label{eq23}
 h(k_1) \longrightarrow\; c(p_1') + \bar{c}(p_2') + \gamma(k_2)\,.
\end{equation}

The calculations for the process~\eqref{eq22}
are straightforward; see appendix~\ref{appA}. The
results for the process
\begin{equation}
\label{eq24}
p(p_1) + \bar{p}(p_2) \longrightarrow\; h(k_1) + \gamma(k_2) + X
\end{equation}
are as follows:
\begin{multline}
\label{eq25}
d \sigma \big( p(p_1) + \bar{p}(p_2) \longrightarrow\; h(k_1) + \gamma(k_2) + X \big) =
\frac{1}{2 s}
\frac{1}{4 N_c^2}
\frac{1}{(2 \pi)^2}\;\\
\times\intop_0^1 d x_1
\intop_0^1 d x_2
\bigg\{
\sum\limits_{q=s,c} \bigg[
\frac{1}{x_1} N_q^p(x_1) 
\frac{1}{x_2} N_{\bar{q}}^{\bar{p}}(x_2) 
I_{q \bar{q}}^h (x_1 p_1, x_2 p_2, k_1, k_2)
+
\frac{1}{x_1} N_{\bar{q}}^p(x_1) 
\frac{1}{x_2} N_{q}^{\bar{p}}(x_2) 
I_{q \bar{q}}^h (x_2 p_2, x_1 p_1, k_1, k_2) \bigg] \bigg\}\\
\times\delta^{(4)}(x_1 p_1+x_2 p_2-k_1-k_2)
\frac{d^3 k_1}{2 k_1^0}
\frac{d^3 k_2}{2 k_2^0}\,.
\end{multline}
The functions $I_{q \bar{q}}^h$ are given
in~\eqref{eqC7} and \eqref{eqC9}.

For the radiative decays
\begin{equation}
\label{eq27}
 h(k_1) \longrightarrow\; q(p_1') + \bar{q}(p_2') + \gamma(k_2)
\end{equation}
with $q=s,c$ we get 
\begin{multline}
\label{eq28}
d \Gamma \big(  h(k_1) \rightarrow\; q(p_1') + \bar{q}(p_2')  + \gamma(k_2) \big) =
\frac{1}{2 m_h} 
\frac{1}{(2 \pi)^5} 
\;\tilde{I}_{q \bar{q}}^h (p_1', p_2', k_1, k_2)
\delta^{(4)}(p_1'+p_2'+k_2-k_1)
\frac{d^3 p_1'}{2 p_1'^0}
\frac{d^3 p_2'}{2 p_2'^0}
\frac{d^3 k_2}{2 k_2^0}\;.
\end{multline}
Here we have from appendix~\ref{appA}
\begin{equation}
\label{eq29}
\tilde{I}_{q \bar{q}}^h (p_1', p_2', k_1, k_2)=
I_{q \bar{q}}^h (p_1', p_2', k_1, -k_2)\,.
\end{equation}

%
\section{Numerical results}
\label{secIV}

\begin{figure*}[t]
\centering
\includegraphics[width=.9\linewidth]{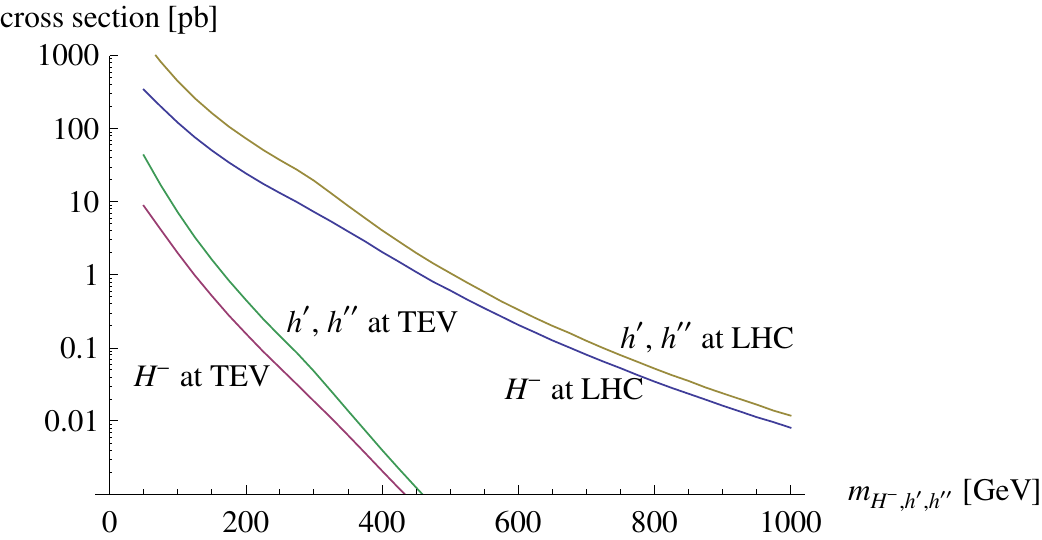}
\caption{\label{fign1} 
Total cross sections for radiative Higgs-boson
production and decay as functions of the Higgs-boson masses.
The results are for Tevatron (lower curves) and LHC (upper curves) energies,
1.96 and 14~TeV, respectively.}
\end{figure*}
In this section we present the numerical results for radiative 
Higgs-boson production and decay at Tevatron and LHC.
As discussed in the introduction
our calculation is performed in the narrow width approximation
for the Higgs bosons.
Then we always have two contributions which 
must be added incoherently: on the one hand we have
Drell--Yan production of a single Higgs boson with its subsequent
radiative decay into a quark pair and a photon. 
On the other hand we have radiative Higgs-boson
production with subsequent
decay of the Higgs boson into a quark pair.

In Fig.~\ref{fign1} we present the 
total cross sections calculated
for the Tevatron~($\sqrt{s} =$~1.96~TeV) and
the LHC~($\sqrt{s} =$~14~TeV) energies as functions
of the Higgs-boson masses.
For $p\bar{p}$ collisions the~$H^+$ and $H^-$
cross sections are equal; for $pp$ collisions they
are practically the same, since the $s$ and $\bar{s}$
as well as the $c$ and $\bar{c}$ pdf's are nearly equal.
In order to avoid collinear and soft regions in
phase space we apply a cut on the
invariant scalar products of 
the relevant momenta~$p$
contracted with
the photon momentum
\begin{equation}
\label{eqmu}
(p\; k_2) \ge \mu\;,
\end{equation}
where $p$ stands for all initial and final state quark
momenta as well as for the momentum of
the Higgs-boson,
$p=p_1',p_2',k_1$.
We fix this minimal cut to~$\mu=100$~GeV$^2$.
Note that in the limit of vanishing 
masses for the s- and c-quarks the cross sections
for radiative $h'$ and $h''$ production
are equal; see appendix~\ref{appA}. With increasing Higgs-boson masses
we find a steep decrease of the total cross
sections over several orders of magnitude.
For the Tevatron we get, due to the much smaller
available phase space, lower cross sections compared
to those for the LHC. Nevertheless,
even for the Tevatron energies the cross sections
are not tiny, that is,
above 0.1~pb for Higgs-boson masses below
200~GeV. At LHC energies the cross sections
are larger than 0.1~pb for Higgs-boson masses up to about 
600~GeV. Of course, the results depend
strongly on the cut in the invariant scalar products,
as discussed below. Note that
the cross sections for the neutral Higgs bosons
are larger than those for $H^-$ by a factor of about two.
This is due to the fact that for $H^-$
production only $s \bar{c}$ annihilation but for
$h'$ and $h''$ both, $s \bar{s}$ and $c \bar{c}$ annihilation,
contribute in proton--proton
(LHC) respectively proton--antiproton (Tevatron)
collisions;
compare \eqref{eq10} with \eqref{eq25}.
But the cross sections for $H^-$ and $H^+$
production taken together are similar to 
the individual cross sections for $h'$ and $h''$.
By a comparison of the
radiative Higgs-boson cross
sections to the Drell--Yan cross sections as discussed
in~\cite{Maniatis:2009vp} we 
find, as expected, the 
radiative cross sections suppressed
by factors of order~$\alpha_{\text{em}}$.

\begin{figure*}[t]
\centering
\includegraphics[width=.9\linewidth]{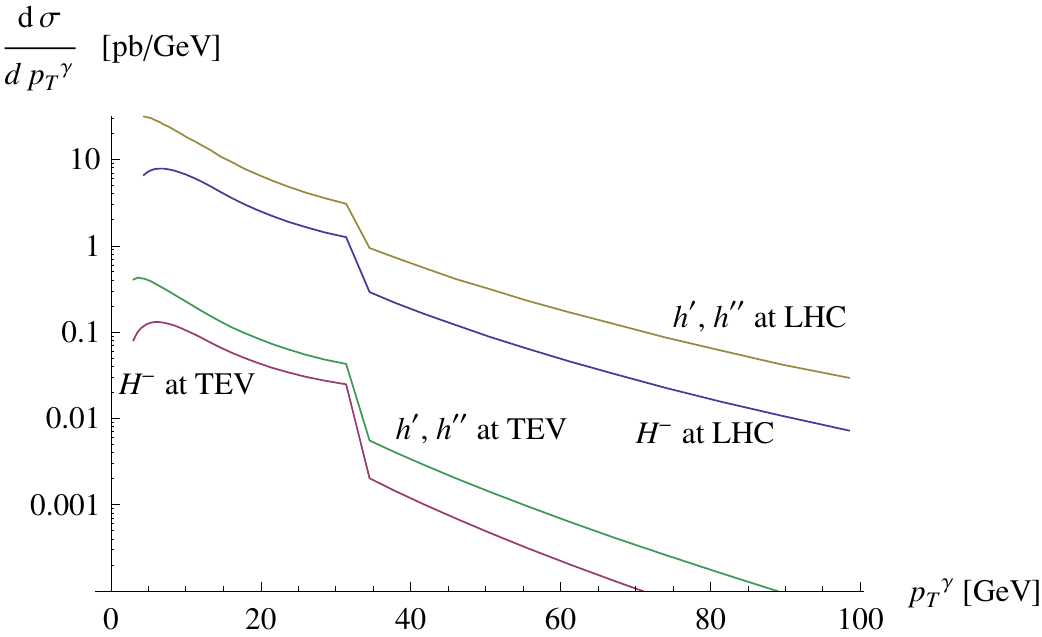}
\caption{\label{fign4} Differential cross section 
$ d \sigma/ d p_\trans^\gamma$ for radiative Higgs-boson
production and decay at LHC (upper pair of curves) and Tevatron energies
(lower pair of curves). 
The Higgs-boson
mass is set to $m_{H^-/h'/h''} = 100$~GeV.}
\end{figure*}
\begin{figure*}[t]
\centering
\includegraphics[width=.9\linewidth]{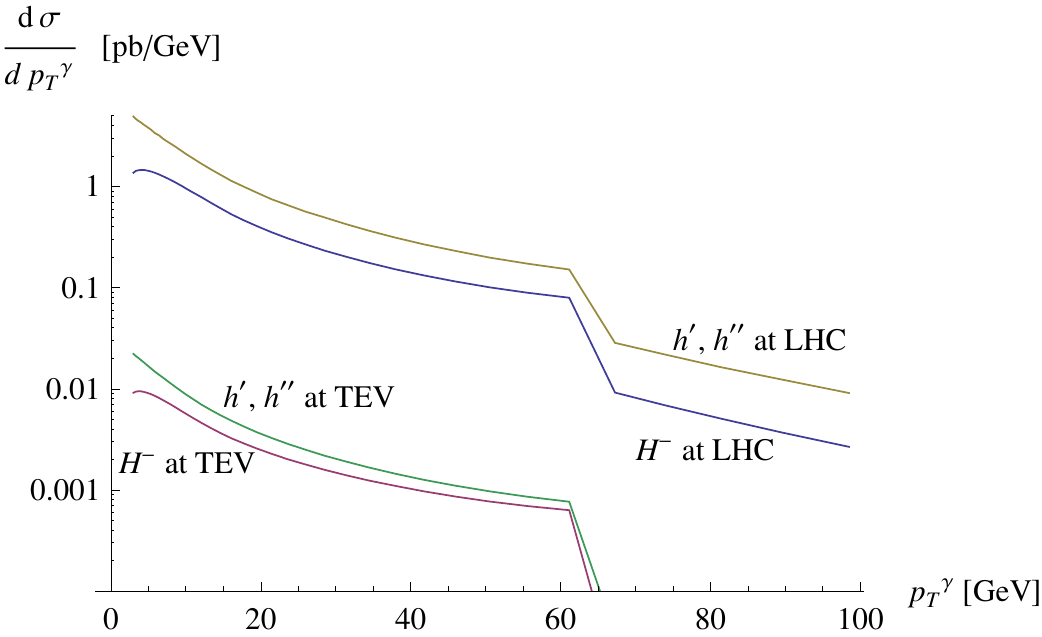}
\caption{\label{fign5} Differential cross section 
$ d \sigma/ d p_\trans^\gamma$ for radiative Higgs-boson
production and decay at LHC (upper pair of curves) and Tevatron energies
(lower pair of curves). 
The Higgs-boson
mass is set to $m_{H^-/h'/h''} = 200$~GeV.}
\end{figure*}
In Figs.~\ref{fign4} and \ref{fign5} we study the
differential cross section
$d \sigma/ d p_\trans^\gamma$ with respect to
the transverse momentum of the photon.
In Fig.~\ref{fign4}
Higgs-boson masses of 100~GeV are assumed
whereas in Fig.~\ref{fign5} we
assume Higgs-boson masses of 200~GeV.
Again a minimal invariant cut~\eqref{eqmu} with $\mu=100$~GeV$^2$ is applied which
leads to the suppression of the
differential cross section for very low
transverse momentum~$p_\trans^\gamma$, that is, below
a few GeV. 
For increasing $p_\trans^\gamma$ we find 
decreasing cross sections, as expected.
An interesting feature of the 
$p_\trans^\gamma$ distributions is
the kink at large $p_\trans^\gamma$ which
is due to
the fact that we have two contributions, 
the radiative Higgs-boson production and
the Drell--Yan single Higgs-boson production
with subsequent radiative decay.
In our calculation we only consider the
lowest order Drell--Yan single Higgs-boson
production and neglect the transverse momenta
of the quarks in the initial proton and antiproton.
Then the transverse momentum of the singly produced
Higgs boson is zero. It is clear that in the subsequent
radiative decays of these Higgs bosons there is
a maximal value for $p_\trans^\gamma$,
essentially half the Higgs-boson mass. Due
to the cuts~\eqref{eqmu} this
kinematic restriction is transformed into
the kink at lower $p_\trans^\gamma$ seen
in Figs.~\ref{fign4} and \ref{fign5}. For values
of $p_\trans^\gamma$ above the kink in essence
only radiative Higgs-boson production, that is,
the processes \eqref{eq5}, \eqref{eq12} and \eqref{eq22}
contribute. In reality the kink will
be washed out for several reasons:
nonzero transverse momenta of the initial
quarks; contributions to the
Drell--Yan process at higher order in the
strong coupling parameter~$\alpha_s$; finite
width effects of the Higgs bosons.
\begin{figure}[h]
\centering
\includegraphics[width=.5\linewidth]{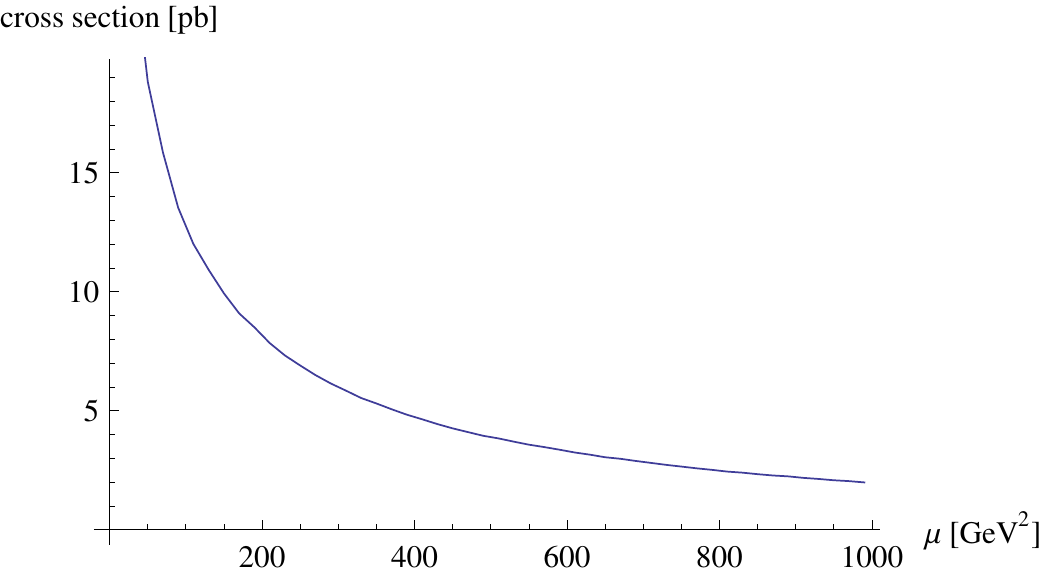}
\caption{\label{fign3} Cross section for the charged Higgs-boson
production $\sigma(p + p \to H^- + \gamma + X)$ at $\sqrt{s}=14$~TeV
as function of the  
the minimal invariant cut $\mu$ defined in~\eqref{eqmu}. 
The charged Higgs-boson mass
is set to $m_{H^-}=200$~GeV. }
\end{figure}

Finally, we 
study in Fig.~\ref{fign3} the dependence of
the cross sections on the invariant 
cut $\mu$~\eqref{eqmu}. 
As an example we show
the radiative production cross section of a charged
Higgs boson via the process $p + p \to H^- + \gamma + X$ 
at $\sqrt{s}=14$~TeV as function of~$\mu$. 
As expected we
get a steeply increasing cross section with
decreasing cut parameter~$\mu$. Of course,
the singularity in the limit of vanishing
$\mu$ is not physical and would be absorbed
by the corresponding virtual photon corrections
to \mbox{$p + p \rightarrow H^- + X$}.
In an experimental analysis of the processes
\eqref{eq1} and \eqref{eq2} energy cuts
and isolation cuts for the photon must 
reflect the real experimental conditions.
Since such cuts are different for each experiment
we have used here only the simple $\mu$ cut~\eqref{eqmu}.
But with the formulas given in the appendix
any cuts can be implemented in a Monte Carlo
program for experimental analysis.

%
\section{Conclusions}
\label{secV}

\begin{figure}
\subfloat{\includegraphics[height=.2\linewidth]{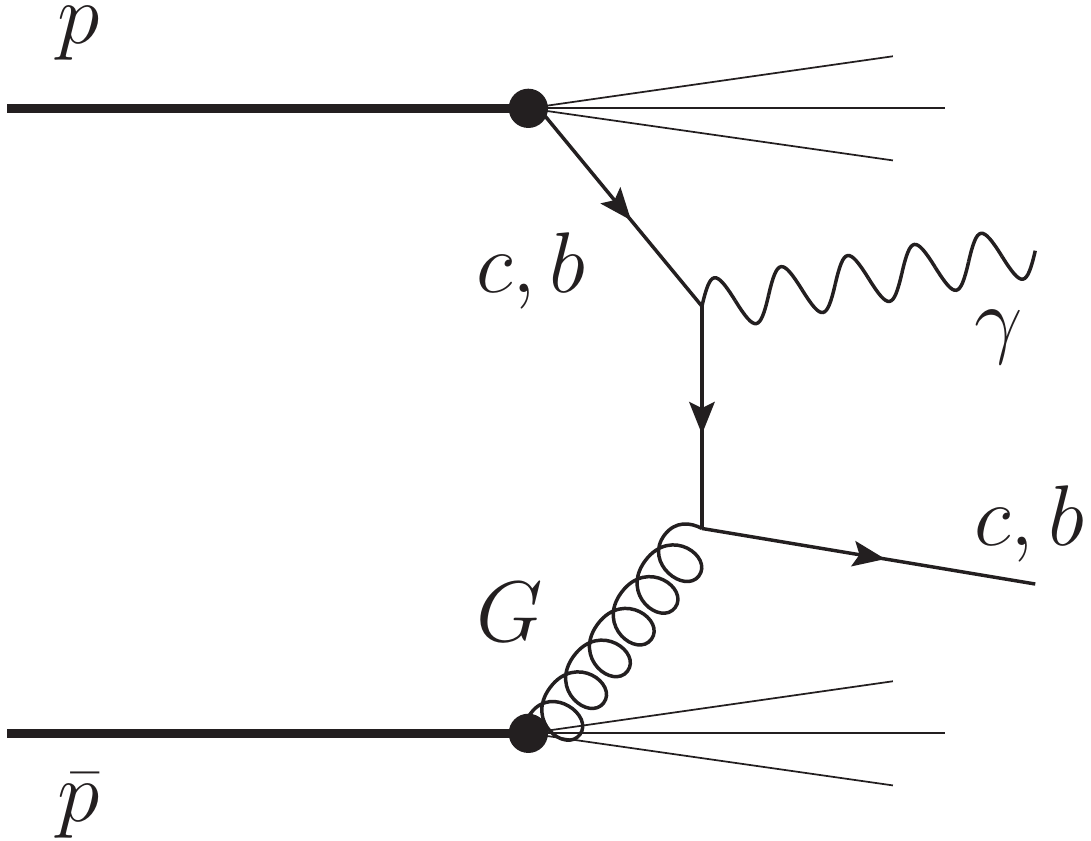}}
\subfloat{\includegraphics[height=.2\linewidth]{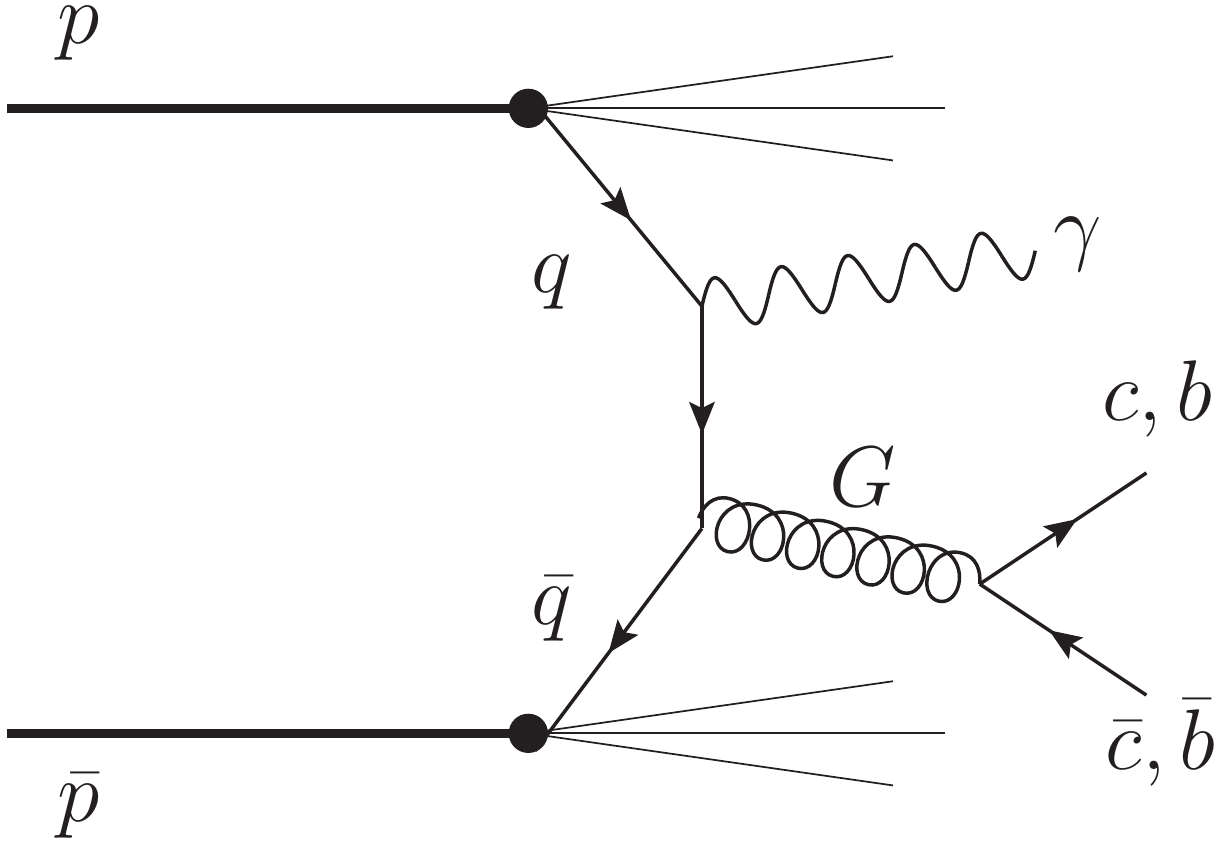}}
\subfloat{\includegraphics[height=.2\linewidth]{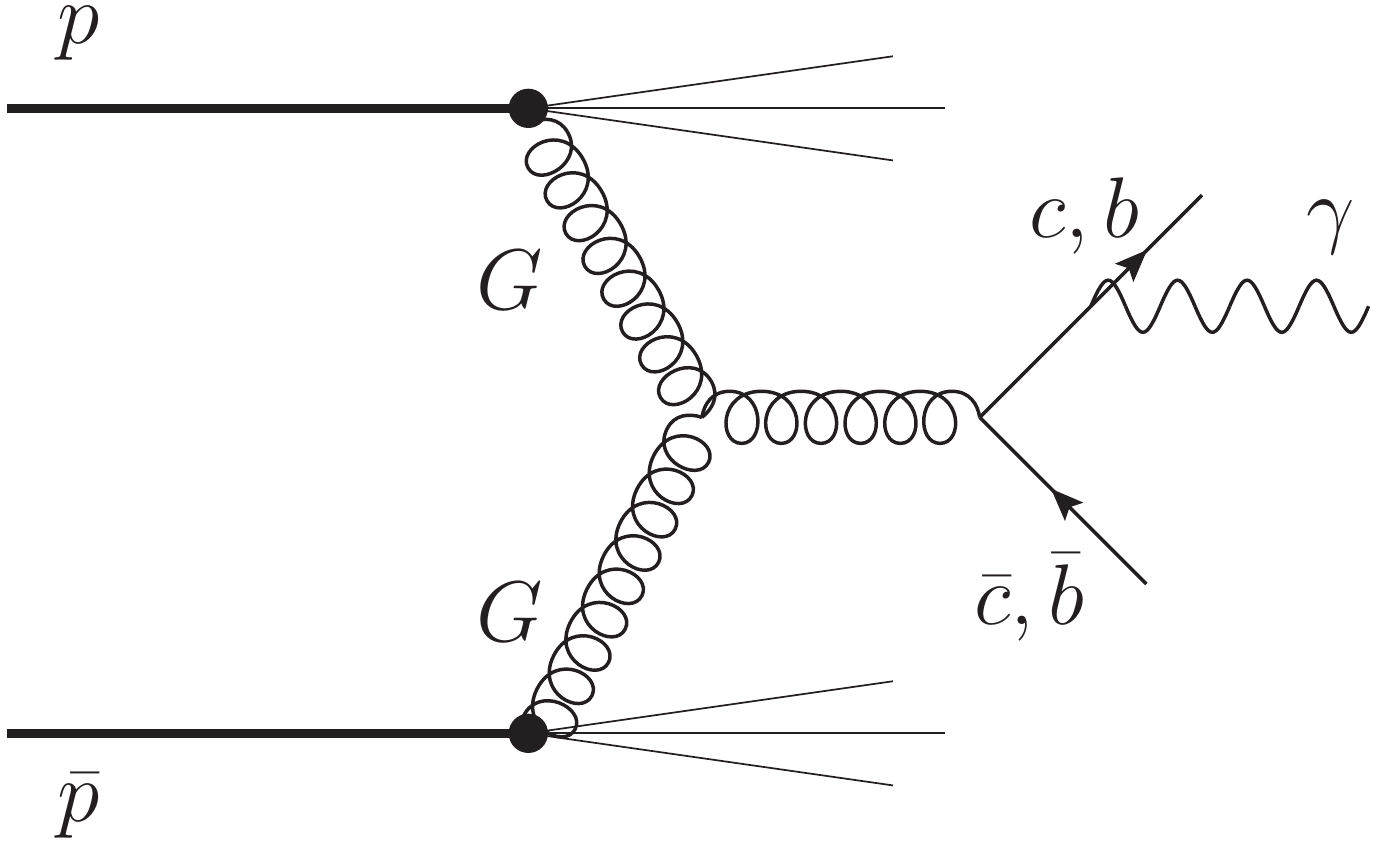}}
\caption{\label{fig9}
Some typical QCD diagrams for processes contributing
to $p + \bar{p} \rightarrow \text{heavy flavour jet(s)}
+ \gamma + X$.}
\end{figure}
In this article we have calculated the simplest processes
contributing to radiative Higgs-boson production and decay
in the MCPM for $pp$ and $p\bar{p}$ collisions at high
energies. As an interesting signal obtainable form these
processes for experiments we discussed the inclusive production
of heavy flavour jet(s) plus a real photon, see \eqref{eq1} and \eqref{eq2}.
In the MCPM only $c$-quark, no $b$-quark jets are produced from
the Higgs channels. Of course, processes \eqref{eq1} and \eqref{eq2}
also occur due to Standard Model QCD effects.
Some diagrams for processes contributing in $p\bar{p}$ collisions are
shown in Fig.~\ref{fig9}.
For $pp$ collisions one just has to replace $\bar{p}$ by $p$.
These QCD processes are a background to the Higgs-boson 
processes (see Figs.~\ref{fig3} and \ref{fig4})
studied in the present paper.

Let us emphasise as distinguishing feature of the MCPM that
we predict only $c$ and $\bar{c}$ quark jets from the Higgs-boson channels.
Thus, the $b$ and $\bar{b}$ jets in the processes
\eqref{eq1} and \eqref{eq2} should show Standard Model QCD behaviour.
In contrast, there should be a surplus of $c$ and $\bar{c}$ jets over
the QCD expectation. We note that due to different final states
there is no interference of the diagrams of Figs.~\ref{fig3} and \ref{fig4}
with the dominant QCD diagrams, examples of which are given in Fig.~\ref{fig9}.
Thus, the QCD and Higgs-boson mediated contributions of the processes
\eqref{eq1} and \eqref{eq2} must be added incoherently.
\begin{figure}
\subfloat[]{\includegraphics[height=.2\linewidth]{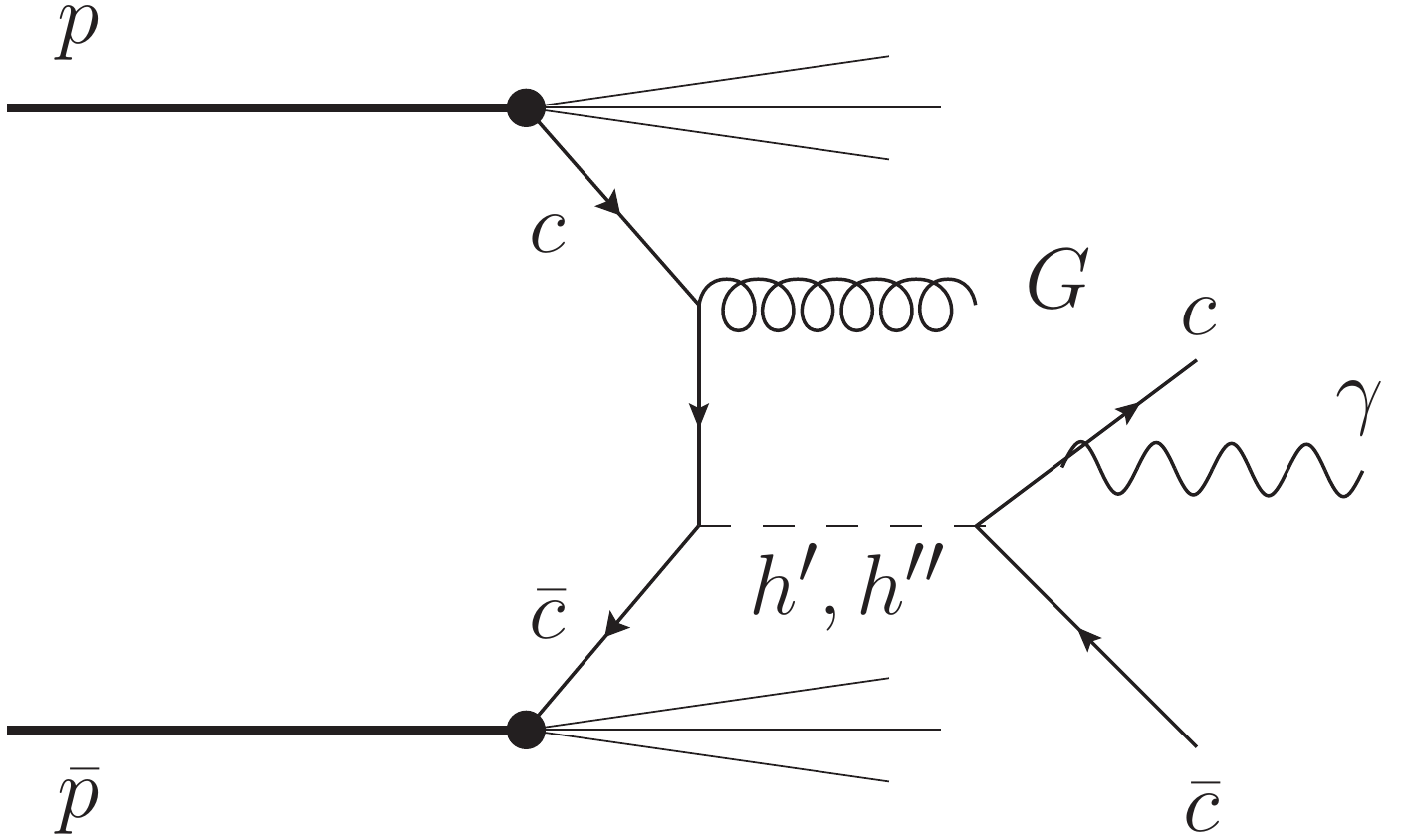}
\label{fig10a}}
\subfloat[]{\includegraphics[height=.2\linewidth]{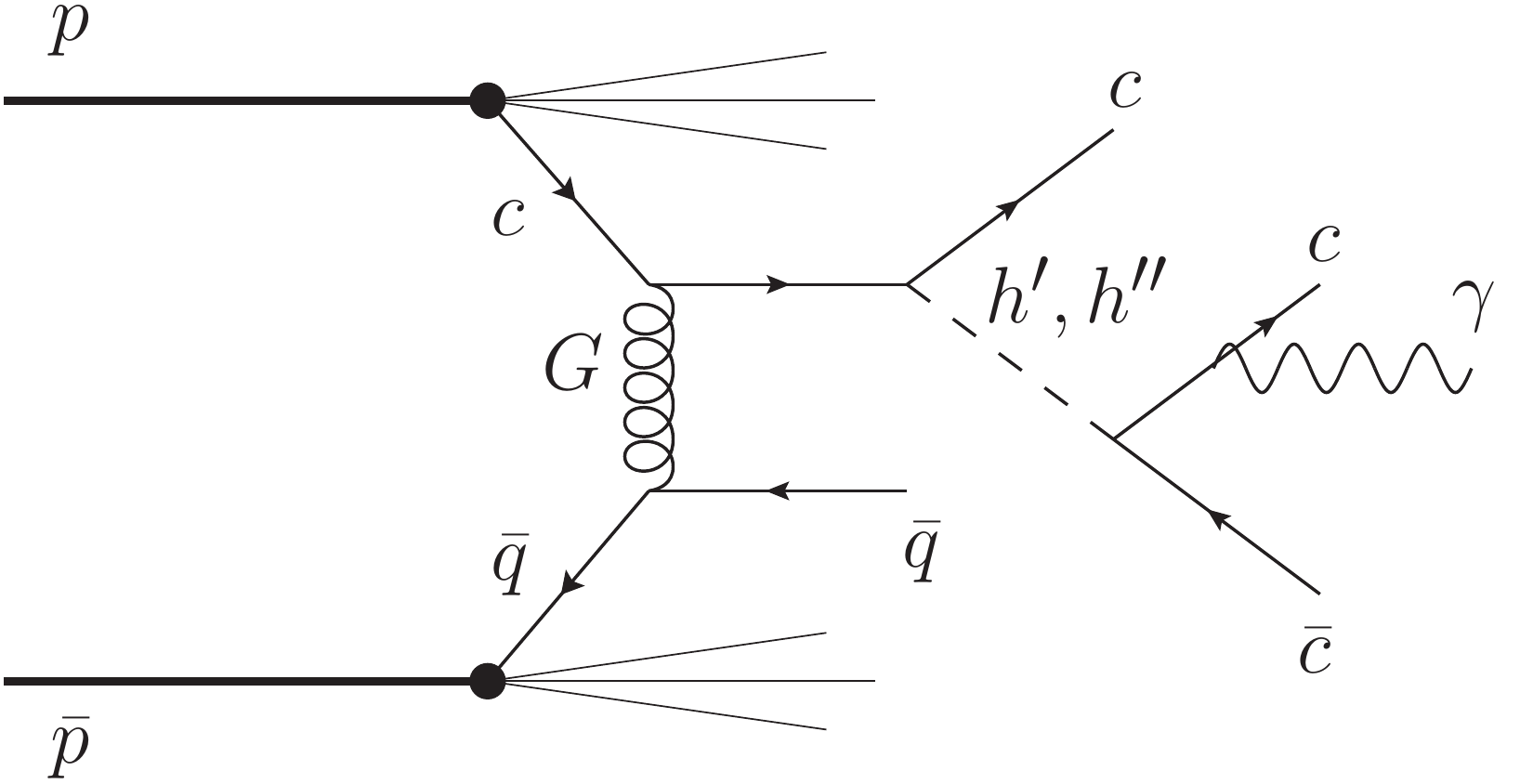}
\label{fig10b}}
\caption{\label{fig10}
(a): Example of a diagram giving a higher order QCD
correction to the Drell--Yan production
of $h'$ and $h''$. 
(b): Example of a diagram where a Higgs boson $h'$ or $h''$
is produced in a $c\bar{q}$ collision where $q$ can be any quark
flavour.
These Higgs bosons may then decay radiatively.
}
\end{figure}

We found cross sections of order 0.1 to 50~pb
for the radiative Higgs-boson production 
at the Tevatron energy $\sqrt{s}=1.96~$TeV. But only for rather light
Higgs bosons the cross section exceeds
1~pb; see Fig.~\ref{fign1}.
In contrast, at $\sqrt{s}=14$~TeV to be reached at the 
LHC we predict large 
cross sections even
for rather heavy Higgs bosons.
For instance for Higgs-boson masses $h',h'',H^{\pm}$
of about 400~GeV we get cross sections exceeding
1~pb. For an integrated luminosity 
of 100~fb$^{-1}$ at LHC this corresponds to at least
100,000 Higgs bosons 
being either produced radiatively or
being produced and then decaying radiatively. 
Even if one order of
magnitude of this number of Higgs bosons is lost due to the separation
of the signal from the background this is still
a sizable number which should be
detectable at the LHC.

Let us note that recently experimental results
for process \eqref{eq1} were published by the D0 collaboration
at Fermilab~\cite{:2009de}. This collaboration observes an
excess over the QCD expectation for the cross section
$\sigma( p \bar{p} \to \gamma + c + X)$
for photons
with a transverse momentum
exceeding 70~GeV.
In contrast, no such excess is observed
for the analogous process 
$\sigma( p \bar{p} \to \gamma + b + X)$.
Qualitatively this finding is as
expected in the MCPM. But what about the
quantitative aspects? First of
all we must mention that the statistical
significance of the excess in the charm
channel is only about 1.6 to 2.2$\sigma$. 
Thus, the effect may well go away with further 
measurements. But, for the sake of argument,
let us take the observed charm jet excess seriously
and try to see if it could have anything to to
with the MCPM. The excess cross section 
$d \sigma/ d p_\trans^\gamma$
which one can deduce from Table~I
of~\cite{:2009de} is of the order of
0.02~pb$/$GeV for $p_\trans^\gamma =$~70
to 90~GeV. 
Comparing with our Figs.~\ref{fign4} and \ref{fign5}
we see that we find cross sections of this level
only for $p_\trans^\gamma < 20$ (10)~GeV for a
Higgs-boson mass of 100~(200)~GeV.
Thus, our calculated cross sections seems
too low compared to the findings of~\cite{:2009de}.
But we have to keep in mind several points before
we can draw definite conclusions.

We used different cuts in our calculations compared
to the experiments. We would invite the
experimentalists to use our formulas as given in
the appendix and to include the processes studied
in this paper into their Monte Carlo programs.
Then all their experimental cuts can be implemented
appropriately.

In the MCPM there are also further mechanisms which will
contribute to the process~\eqref{eq1}.
We have higher order QCD corrections to the
Drell--Yan production of the Higgs bosons,
see Fig.~\ref{fig10a}. These processes are suppressed
by a factor $\alpha_s$ relative to the leading
order Drell--Yan process. But, clearly, diagrams
as shown in Fig.~\ref{fig10a} will lead to Higgs bosons
produced at substantial values of $p_\trans$. Then,
the radiative decay of these Higgs bosons will lead to
photons of much higher $p_\trans^\gamma$ than from
the leading order process.
Furthermore there is the possibility of a sort of Higgs-strahlung
process in $p\bar{p}$ collisions. An example in shown 
in Fig.~\ref{fig10b}. Since the coupling of the $c$ quark
to $h'$ and $h''$ is of order 1 the rates for such processes could
be quite large. Clearly, for such processes there are 
the analogous radiative production ones which will 
also contribute to~\eqref{eq1}.\\

In summary we can say the following concerning the findings
of~\cite{:2009de}.
If the charm excess seen is real and confirmed by further
experiments it is not excluded that it may have 
something to do with Higgs-boson production in the MCPM.
Our calculations of the cross sections and the
$p_\trans^\gamma$ distributions for the 
processes~\eqref{eq1} and \eqref{eq2} must be considered
as giving only lower limits for these quantities in the MCPM.
We have identified above various channels which will also
contribute but still have to be calculated theoretically.
Nonetheless we can say that - at least to us - the magnitude
of the effects calculated so far looks interesting for the
Tevatron experiments. Thus, a study of the invariant mass
distributions of the channels $c\bar{c}$, $c\bar{s}$, $s\bar{c}$,
$c\bar{c}\gamma$, $c\bar{s}\gamma$, and $s\bar{c}\gamma$ could
be interesting.\\

We hope that our remarks will be useful to experimentalists and will
induce them to study further heavy flavour jet(s) plus photon
inclusive production in $p\bar{p}$ and $pp$ collisions
at the Tevatron and the LHC, respectively.

\acknowledgments{
We thank H.~C.~Schultz-Coulon for useful discussions. This work
was supported by Deutsche Forschungsgemeinschaft, project number 
\mbox{NA 296/5-1}.
}

\appendix
\numberwithin{equation}{section}

%
\section{}
\label{appA}

Here we give some details of the calculation for the
radiative production
and radiative decays of~$h', h''$ and $H^\pm$.\\

Let us first consider the reaction~\eqref{eq5} and
the corresponding T-matrix element for production
of a photon with polarisation vector~$\epsilon$
\begin{equation}
\label{eqa100}
\epsilon^{* \mu} {\cal T}_\mu^{H^-} =
\langle H^-(k_1), \gamma(k_2, \epsilon)\; |\; T\; |\; s(p_1'), \bar{c}(p_2') \rangle\;.
\end{equation}
Here and in the following the colour and spin indices of the quarks
are not written out explicitly. The function $I_{s\bar{c}}^{H^-}$
is defined as
\begin{equation}
\label{eqa101}
I_{s\bar{c}}^{H^-} (p_1', p_2', k_1, k_2) =
\sum_{\text{spins, colours}} (-1)\;{\cal T}_\mu^{H^-}({\cal T}^{H^-\mu})^*\;.
\end{equation}
The calculation of this squared amplitude for the
process~\eqref{eq5} is straightforward and yields 
\begin{multline}
\label{eqA6}
I_{s \bar{c}}^{H^-} (p_1',p_2', k_1,k_2) =
\frac{4 e^2}{v_0^2} N_c
  \bigg\{\\
  \frac{Q_s^2}{(p_1' k_2)^2} 
    \bigg[
    (m_t^2+m_b^2) ((p_1' k_2)(p_2'k_2)
    -(p_1' p_2') m_s^2 + (p_2' k_2) m_s^2)
    + 2 m_t m_b m_c m_s ((p_1' k_2) -m_s^2) 
    \bigg] + \\
  \frac{Q_c^2}{(p_2' k_2)^2} 
    \bigg[
    (m_t^2+m_b^2) ((p_1' k_2)(p_2'k_2)
    -(p_1' p_2') m_c^2 + (p_1' k_2) m_c^2)
    + 2 m_t m_b m_c m_s ((p_2' k_2) -m_c^2) 
    \bigg] +\\
  \frac{1}{(k_1 k_2)^2}
    \bigg[
    -(m_t^2+ m_b^2)(p_1' p_2')(m_{H^-}^2+(p_1'k_2) + (p_2'k_2))
    -2 m_t m_b m_c m_s (m_{H^-}^2 +(p_1' k_2)+(p_2' k_2))
    \bigg] +\\
\frac{Q_s Q_c}{(p_1' k_2) (p_2' k_2)}
    \bigg[
    (m_t^2+m_b^2) ( 2 (p_1' p_2')^2 - 2(p_1' p_2')(p_1' k_2)
    - 2(p_1' p_2')(p_2' k_2)
    +2 (p_1' k_2)(p_2' k_2) +\\ 
    m_c^2 (p_1' k_2) + m_s^2 (p_2' k_2))
    + 2 m_t m_b m_c m_s ( 2 (p_1' p_2') - (p_1' k_2) - (p_2' k_2))
    \bigg] +\\
  \frac{Q_c}{(p_2' k_2) (k_1 k_2)}
    \bigg[
    (m_t^2+m_b^2) ( 2 (p_1' p_2')^2 - 2(p_1' p_2')(p_1' k_2)
     - (p_1' p_2')(p_2' k_2)\\
    +2 m_c^2 (p_1' p_2') - m_c^2 (p_1' k_2) + m_s^2 (p_2' k_2))
    + 2 m_t m_b m_c m_s ( 2 (p_1' p_2') - (p_1' k_2) - 2(p_2' k_2) + 2 m_c^2)
    \bigg] +\\
  \frac{Q_s}{(p_1' k_2) (k_1 k_2)}
    \bigg[-
    (m_t^2+m_b^2) ( 2 (p_1' p_2')^2 - 2(p_1' p_2')(p_2' k_2)
     - (p_1' p_2')(p_1' k_2)\\
    +2 m_s^2 (p_1' p_2') + m_c^2 (p_1' k_2) - m_s^2 (p_2' k_2))
    + 2 m_t m_b m_c m_s ( -2 (p_1' p_2') + 2(p_1' k_2) + (p_2' k_2) - 2 m_s^2)
    \bigg] 
  \bigg\}\;.
\end{multline}
Here $Q_s=-1/3$ and $Q_c=2/3$ are the charges of the $s$ and $c$ quark
in units of the positron charge, respectively.
The standard Higgs-boson vacuum
expectation value is $v_0 \approx 246$~GeV. 
The function $I_{s \bar{c}}^{H^-}$ is
defined only for momenta satisfying
energy-momentum conservation, $p_1'+p_2'=k_1+k_2$.

We can always use energy-momentum conservation to eliminate one
of the momenta in~\eqref{eqA6}. Since the momentum
one may want to eliminate differs from case to case
we keep the more symmetric but redundant notation
of~\eqref{eqA6}.
We have kept the masses
of all quarks non-vanishing. Of course
the expression~\eqref{eqA6} simplifies
considerably
if we neglect the s- and c-quark masses.\\


Next we consider the reaction~\eqref{eq12}.
We can use the standard CP transformation
to relate the amplitudes for the 
reactions~\eqref{eq5} and \eqref{eq12}. 
This transformation, denoted by $\text{CP}_s \equiv \CPb$
in~\cite{Maniatis:2007de,Maniatis:2009vp},
is conserved in the MCPM
(see sect.~3 of~\cite{Maniatis:2007de})
and therefore
we find
\begin{equation}
\label{eqa12}
\epsilon^{* \mu} {\cal T}_\mu^{H^+} =
\langle H^+(k_1), \gamma(k_2,\epsilon)\; |\; T\; |\; \bar{s}(p_1'), c(p_2') \rangle
=
\langle H^-(\tilde{k}_1), \gamma(\tilde{k}_2, \tilde{\epsilon})\; |\; T\; |\; s(\tilde{p}_1'), \bar{c}(\tilde{p}_2') \rangle
\end{equation}
where
\begin{equation}
\tilde{k}_{1,2}^\lambda = 
k_{1,2\; \lambda}\;, \qquad
\tilde{p}_{1,2}^{'\lambda} = 
p'_{1,2\; \lambda}\;, \qquad
\tilde{\epsilon}^\lambda = 
\epsilon_\lambda\;.
\end{equation}
Analogously to~\eqref{eqa101} we define
\begin{equation}
I_{\bar{s}c}^{H^+} (p_1', p_2', k_1, k_2) =
\sum_{\text{spins, colours}} (-1)\; {\cal T}_\mu^{H^+}({\cal T}^{H^+\mu})^*\;.
\end{equation}
and find from~\eqref{eqa101} and \eqref{eqa12}
\begin{equation}
\label{eqa15}
I_{\bar{s}c}^{H^+} (p_1', p_2', k_1, k_2) =
I_{s\bar{c}}^{H^-} (\tilde{p}_1', \tilde{p}_2', \tilde{k}_1, \tilde{k}_2)\;.
\end{equation}
Since $I_{s\bar{c}}^{H^-}$ depends only on three linearly independent
four momenta there is no parity odd scalar which can be formed. Thus,
we get
\begin{equation}
\label{eqa16}
I_{s\bar{c}}^{H^-}(\tilde{p}_1', \tilde{p}_2', \tilde{k}_1, \tilde{k}_2) =
I_{s\bar{c}}^{H^-}(p_1', p_2', k_1, k_2)\;,
\end{equation}
as can also be verified directly form~\eqref{eqA6}.
Combining~\eqref{eqa15} and \eqref{eqa16} we obtain
\begin{equation}
\label{eqa17}
I_{\bar{s}c}^{H^+}(p_1', p_2', k_1, k_2) =
I_{s\bar{c}}^{H^-}(p_1', p_2', k_1, k_2)\;.
\end{equation}

For the reaction~\eqref{eq22} we write the 
T-matrix element similarly to~\eqref{eqa100}
\begin{equation}
\label{eqa200}
\epsilon^{* \mu}{\cal T}_{q\bar{q},\;\mu}^{h} =
\langle h(k_1), \gamma(k_2,\epsilon) \;|\; T\; |\; q(p_1'), \bar{q}(p_2') \rangle
\end{equation}
with $q=s, c$. The corresponding function $I_{q\bar{q}}^h$ is defined as
\begin{equation}
I_{q\bar{q}}^h (p_1', p_2', k_1, k_2) =
\sum_{\text{spins, colours}} (-1)\; {\cal T}_{q\bar{q},\;\mu}^h
({\cal T}_{q\bar{q}}^{h\;\mu})^* \;.
\end{equation}
This squared amplitude reads for $h=h'$ 
\begin{multline}
\label{eqC7}
I_{s \bar{s}/ c \bar{c}}^{h'} (p_1',p_2', k_1,k_2) =
\left(\frac{Q_{s/c}\; e}{v_0}\right)^2 4 N_c m_{b/t}^2
  \bigg\{\\
 \frac{2}{(p_1' k_2) (p_2' k_2)}
    \bigg[
    (p_1' p_2')^2 - (p_1' p_2')(p_1' k_2)
    - (p_1' p_2')(p_2' k_2)
    + m_{s/c}^2 (p_1' k_2)
    + m_{s/c}^2 (p_2' k_2)
    - m_{s/c}^2 (p_1' p_2')
    \bigg]\\
  -\frac{m_{s/c}^2}{(p_1' k_2)^2} 
    \bigg[
    (p_1' p_2') + (p_1' k_2)
    -(p_2' k_2) -m_{s/c}^2
    \bigg]
  -\frac{m_{s/c}^2}{(p_2' k_2)^2} 
    \bigg[
    (p_1' p_2') - (p_1' k_2)
    +(p_2' k_2) -m_{s/c}^2
    \bigg]
  +\frac{1}{(p_1' k_2) (p_2' k_2)}
    \bigg[
    (p_1' k_2)+(p_2' k_2)
    \bigg]^2
  \bigg\}\,.
\end{multline}
For $h=h''$ we find
\begin{multline}
\label{eqC9}
I_{s \bar{s}/ c \bar{c}}^{h''} (p_1',p_2', k_1,k_2) =
\left(\frac{Q_{s/c} e}{v_0}\right)^2 4 N_c m_{b/t}^2
  \bigg\{\\
    \bigg[
    \frac{2(p_1' p_2')}{(p_1' k_2) (p_2' k_2)}
    - \frac{m_{s/c}^2}{(p_1' k_2)^2}
    - \frac{m_{s/c}^2}{(p_2' k_2)^2}
    \bigg]\; 
    \bigg[
    (p_1' p_2') - (p_1' k_2)
    -(p_2' k_2) + m_{s/c}^2
    \bigg]
  +\frac{1}{(p_1' k_2) (p_2' k_2)}
    \bigg[
    (p_1' k_2)+(p_2' k_2)    
    \bigg]^2
  \bigg\}.
\end{multline}
We see that for vanishing masses $m_s=m_c=0$ we have
$I_{s \bar{s}/ c \bar{c}}^{h''}=I_{s \bar{s}/ c \bar{c}}^{h'}$.\\

Let us now discuss the radiative decays of $H^-$~\eqref{eq17}
and $H^+$~\eqref{eq18}.
We define the corresponding amplitudes as
\begin{equation}
\label{eqa300}
\epsilon^{* \mu}\tilde{{\cal T}}_\mu^{H^-} =
\langle s(p_1'), \bar{c}(p_2') , \gamma(k_2,\epsilon) \;|\; T\; |\; H^-(k_1)  \rangle,
\qquad
\epsilon^{* \mu}\tilde{{\cal T}}_\mu^{H^+} =
\langle \bar{s}(p_1'), c(p_2') , \gamma(k_2,\epsilon) \;|\; T\; |\; H^+(k_1)  \rangle\;.
\end{equation}
Next we define
\begin{equation}
\tilde{I}_{s\bar{c}}^{H^-} (p_1', p_2', k_1, k_2) =
\sum_{\text{spins, colours}} (-1)\; \tilde{{\cal T}}_\mu^{H^-}(\tilde{{\cal T}}^{H^-\mu})^*\;,
\quad
\tilde{I}_{\bar{s}c}^{H^+} (p_1', p_2', k_1, k_2) =
\sum_{\text{spins, colours}} (-1)\; \tilde{{\cal T}}_\mu^{H^+}(\tilde{{\cal T}}^{H^+\mu})^*\;.
\end{equation}
It is easy to see that the following crossing relation holds
\begin{equation}
\label{eqa303}
\tilde{{\cal T}}_\mu^{H^-}= ({\cal T}_\mu^{H^-})^* \big|_{k_2 \to - k_2}\;,
\end{equation}
where ${\cal T}_\mu^{H^-}$ is defined in~\eqref{eqa100}.
From~\eqref{eqa303} and \eqref{eqa101} we get immediately
\begin{equation}
\label{eqa304}
\tilde{I}_{s\bar{c}}^{H^-}({p}_1', {p}_2', {k}_1, {k}_2) =
I_{s\bar{c}}^{H^-}(p_1', p_2', k_1, - k_2)\;,
\end{equation}
where $k_1=p_1'+p_2'+k_2$. The same CP arguments as used in~\eqref{eqa12}
to \eqref{eqa17} give also
\begin{equation}
\label{eqa305}
\tilde{I}_{\bar{s}c}^{H^+}(p_1', p_2', k_1, k_2) =
I_{s\bar{c}}^{H^-}(p_1', p_2', k_1, - k_2)\;.
\end{equation}

Finally we discuss the decay
\begin{equation}
h(k_1) \rightarrow q(p_1') + \bar{q}(p_2') +\gamma(k_2,\epsilon)\;,
\end{equation}
with $q=s,\;c$ and $h=h',\; h''$.
The corresponding amplitude and the squared amplitude are defined
as follows
\begin{equation}
\label{eqa307}
\epsilon^{* \mu} \tilde{{\cal T}}_{q \bar{q},\;\mu}^h =
\langle q(p_1'), \bar{q}(p_2') , \gamma(k_2,\epsilon) \;|\; T\; |\; h(k_1)  \rangle\;,
\end{equation}
\begin{equation}
\label{eqa308}
\tilde{I}_{q \bar{q}}^h(p_1', p_2', k_1, k_2) =
\sum_{\text{spins, colours}} (-1)\; \tilde{{\cal T}}_{q \bar{q},\;\mu}^h
(\tilde{{\cal T}}_{q \bar{q}}^{h\;\mu})^*\;,
\end{equation}
With the same crossing arguments as used to derive~\eqref{eqa303}
and \eqref{eqa304} we find
\begin{equation}
\label{eqa309}
\tilde{I}_{q\bar{q}}^h({p}_1', {p}_2', {k}_1, {k}_2) =
I_{q\bar{q}}^h(p_1', p_2', k_1, - k_2)\;,
\end{equation}
where $k_1=p_1'+p_2'+k_2$.

%
%
%


\begin{thebibliography}{99}



\bibitem{Kobayashi:1973fv}
  M.~Kobayashi and T.~Maskawa,
  \mbox{Prog.\ Theor.\ Phys.\  {\bf 49}} 652 (1973)

\bibitem{Lee:1973iz}
  T.~D.~Lee,
  Phys.\ Rev.\  D {\bf 8} 1226 (1973).

\bibitem{Lee:1974jb}
  T.~D.~Lee,
  Phys.\ Rept.\  {\bf 9} 143 (1974) .

\bibitem{Glashow:1976nt}
  S.~L.~Glashow and S.~Weinberg,
  Phys.\ Rev.\  D {\bf 15} 1958 (1977).

\bibitem{Paschos:1976ay}
  E.~A.~Paschos,
  Phys.\ Rev.\  D {\bf 15} 1966 (1977).



\bibitem{Gunion:1989we}
  J.~F.~Gunion, H.~E.~Haber, G.~L.~Kane and S.~Dawson,
  ``The Higgs Hunter's Guide'', 
  Addison--Wesley publishing company (1990).

\bibitem{Cvetic:1993cy}
  G.~Cvetic,
  \mbox{Phys.\ Rev.\  D {\bf 48}} 5280 (1993)
  \mbox{[hep-ph/9309202]}.


\bibitem{Ginzburg:2004vp}
  I.~F.~Ginzburg and M.~Krawczyk,
  \mbox{Phys.\ Rev.\ D {\bf 72}} 115013 (2005)
  \mbox{[hep-ph/0408011]}.


\bibitem{Gunion:2005ja}
J.~F.~Gunion and H.~E.~Haber,
\mbox{Phys.\ Rev.\ D {\bf 72}} 095002 (2005) 
\mbox{[hep-ph/0506227v2]}.


\bibitem{Barbieri:2005kf}
  R.~Barbieri and L.~J.~Hall,
  ``Improved naturalness and the two Higgs doublet model'',
  \mbox{[hep-ph/0510243]}.


\bibitem{Branco:2005em}
  G.~C.~Branco, M.~N.~Rebelo and J.~I.~Silva-Marcos,
  \mbox{Phys.\ Lett.\  B {\bf 614}} 187 (2005)
  \mbox{[hep-ph/0502118]}.

\bibitem{Barbieri:2006dq}
  R.~Barbieri, L.~J.~Hall and V.~S.~Rychkov,
  \mbox{Phys.\ Rev.\  D {\bf 74}} 015007 (2006)
  \mbox{[hep-ph/0603188]}.

\bibitem{Nishi:2006tg}
  C.~C.~Nishi,
  Phys.\ Rev.\  D {\bf 74}, 036003 (2006)
  [Erratum-ibid.\  D {\bf 76} 119901 (2007)]
  [arXiv:hep-ph/0605153].

\bibitem{Ivanov:2006yq}
  I.~P.~Ivanov,
  Phys.\ Rev.\  D {\bf 75}, 035001 (2007)
  [Erratum-ibid.\  D {\bf 76} 039902 (2007)]
  [arXiv:hep-ph/0609018].


\bibitem{Fromme:2006cm}
  L.~Fromme, S.~J.~Huber and M.~Seniuch,
  \mbox{JHEP {\bf 0611}} 038 (2006)
  \mbox{[hep-ph/0605242]}.

\bibitem{Barroso:2007rr}
  A.~Barroso, P.~M.~Ferreira and R.~Santos,
  Phys.\ Lett.\  B {\bf 652} 181 (2007)
  [arXiv:hep-ph/0702098].


\bibitem{Gerard:2007kn}
  J.~M.~Gerard and M.~Herquet,
  \mbox{Phys.\ Rev.\ Lett.\  {\bf 98}} 251802 (2007)
  \mbox{[hep-ph/0703051]}.


\bibitem{Ma:2009ax}
  E.~Ma and M.~Maniatis,
  arXiv:0909.2855 [hep-ph].
  
\bibitem{Ferreira:2009wh}
  P.~M.~Ferreira, H.~E.~Haber and J.~P.~Silva,
  Phys.\ Rev.\  D {\bf 79} 116004 (2009)
  [arXiv:0902.1537 [hep-ph]].


\bibitem{Mahmoudi:2009zx}
  F.~Mahmoudi and O.~Stal,
  arXiv:0907.1791 [hep-ph].



\bibitem{Bernreuther:1998rx}
  W.~Bernreuther and O.~Nachtmann,
  Eur.\ Phys.\ J.\  C {\bf 9} 319 (1999)
  [arXiv:hep-ph/9812259].
  


\bibitem{Nagel:2004sw}
  F.~Nagel,
  ``New aspects of gauge-boson couplings and the Higgs sector,'' (2004)
\href{http://www.slac.stanford.edu/spires/find/hep/www?irn=6461018}{SPIRES entry}



\bibitem{Maniatis:2006fs}
  M.~Maniatis, A.~von Manteuffel, O.~Nachtmann and F.~Nagel,
  \mbox{Eur.\ Phys.\ J.\  C {\bf 48}} 805 (2006)
  \mbox{[hep-ph/0605184]}.


\bibitem{Maniatis:2006jd}
  M.~Maniatis, A.~von Manteuffel and O.~Nachtmann,
  \mbox{Eur.\ Phys.\ J.\  C {\bf 49}} 1067 (2007)
  [arXiv:hep-ph/0608314].
  
\bibitem{Maniatis:2007vn}
  M.~Maniatis, A.~von Manteuffel and O.~Nachtmann,
  \mbox{Eur.\ Phys.\ J.\  C {\bf 57}} 719 (2008)
  \mbox{[arXiv:0707.3344 [hep-ph]]}.

\bibitem{Maniatis:2007de}
  M.~Maniatis, A.~von Manteuffel and O.~Nachtmann,
  \mbox{Eur.\ Phys.\ J.\  C {\bf 57}} 739 (2008)
  \mbox{[arXiv:0711.3760 [hep-ph]]}.


\bibitem{Maniatis:2009vp}
  M.~Maniatis and O.~Nachtmann,
  \mbox{JHEP {\bf 0905}} 028 (2009)
  [arXiv:0901.4341 [hep-ph]].


\bibitem{:2009de}
  V.~M.~Abazov {\it et al.}  [D0 Collaboration],
  \mbox{Phys.\ Rev.\ Lett.\  {\bf 102} } 192002 (2009)
  [arXiv:0901.0739 [hep-ex]].

  

\bibitem{Nachtmann:1990ta}
  O.~Nachtmann,
  ``Elementary Particle Physics: Concepts And Phenomena'',
  Springer, Berlin (1990).



\end{thebibliography}
\end{document}